\documentstyle[aps,epsf,floats]{revtex}
\begin{document}
\title{Non-perturbative approach to the attractive Hubbard model}
\author{S. Allen, and A.-M.S. Tremblay$^{1}$}
\address{D\'{e}partement de Physique and Centre de recherche sur les\\
propri\'{e}t\'{e}s \'{e}lectroniques de mat\'{e}riaux avanc\'{e}s.\\
$^{1}$Institut canadien de recherches avanc\'{e}es}
\address{Universit\'{e} de Sherbrooke, Sherbrooke, Qu\'{e}bec, Canada J1K 2R1.}
\date{\today}
\maketitle

\begin{abstract}
A non-perturbative approach to the single-band attractive Hubbard model is
presented in the general context of functional derivative approaches to
many-body theories. As in previous work on the repulsive model, the first
step is based on a local-field type ansatz, on enforcement of the Pauli
principle and a number of crucial sum-rules. The Mermin-Wagner theorem in
two dimensions is automatically satisfied. At this level, two-particle
self-consistency has been achieved. In the second step of the approximation,
an improved expression for the self-energy is obtained by using the results
of the first step in an exact expression for the self-energy where the high-
and low-frequency behaviors appear separately. The result is a cooperon-like
formula. The required vertex corrections are included in this self-energy
expression, as required by the absence of a Migdal theorem for this problem.
Other approaches to the attractive Hubbard model are critically compared.
Physical consequences of the present approach and agreement with Monte Carlo
simulations are demonstrated in the accompanying paper (following this one).

\end{abstract}


\section{Introduction}

In the mid to late 1950's, quantum-field theoretical methods that had been
developed first in the context of quantum electrodynamics began to have
widespread applications in Condensed Matter Physics.\cite
{MartinSchwinger:1959,Pines:1962} One can, roughly speaking, distinguish two
types of approaches. The diagrammatic methods of Feynman and the functional
methods of the Schwinger school\cite{KadanoffBaym:1962,Baym:1962}. Both
points of view on many-body theory are equivalent. In particular,
perturbation theory can be formulated diagrammatically or with functional
methods. The two approaches in fact complement each other. For example, in
calculating response functions, subsets of diagrams are often summed to
infinite order. But naive resummations will generally break gauge invariance
or other exact symmetry properties unless consistency between self-energy
and two-particle irreducible vertices is enforced following a technique
whose most natural formulation employs functional derivatives\cite
{Baym:1962,Bickers:1989}.

Diagrammatic methods have nevertheless become by far the most popular
techniques for many-body\cite{Ncorps} problems in Condensed Matter, but the
quest for non-perturbative approaches leads in general outside the realm of
diagrams. While the Hartree-Fock approximation has a diagrammatic
interpretation, what seems to be the most accurate approach to the electron
gas at metallic densities (local field approximation (LFA)\cite{Singwi:1968}%
) does not have a simple diagrammatic interpretation. In the case of the
Hartree-Fock approximation, a variational principle guides the accuracy of
the approximation. In the LFA, it is a self-consistency requirement at the
two particle level that controls the accuracy.

Non-perturbative approaches have been developed in particular for the
Hubbard model\cite{Hubbard:1963}, perhaps the best know model for strongly
interacting electrons on a lattice. While the early Green function
decoupling schemes have largely fallen in disfavor because of their {\it ad
hoc} and uncontrolled nature, many successful non-perturbative approaches
have been developed in one dimension. Recently, dynamical mean-field theory
has become the method of choice\cite{Georges:1996} for higher dimensions.
This method, however, is in fact based on an expansion about infinite
dimension. In two dimensions in particular, the momentum dependence of the
self-energy cannot be neglected and this method becomes less accurate.

The purpose of this paper is to extend to the attractive Hubbard model a
non-perturbative approach developed previously for the repulsive model\cite
{Vilk:1994,Vilk:1995,Vilk:1997}. This method was an extension of the LFA
work of Singwi, Tosi, Land and Sj\"{o}lander\cite{Singwi:1968} for the
electron gas and of Hedeyati and Vignale for the Hubbard model\cite
{Vignale:1989}. It went further than these work in imposing the Pauli
principle, a number of exact sum rules, of conservation laws, and proposing
a formula for the self-energy in the paramagnetic state that includes
momentum and frequency dependence. The approach also includes an internal
check on accuracy based on an exact relationship between one and
two-particle properties. Although the method fails in strong coupling or
very close to a critical point, it gives the most accurate results when
compared with Monte Carlo simulations in the weak to intermediate coupling
regime\cite{Vilk:1997}. The present paper generalizes that approach to the
attractive Hubbard model. The accompanying paper\cite{KyungAllen:2000}
demonstrates accuracy by comparisons with Monte Carlo simulations, and
discusses a problem of importance in the context of high-temperature\cite
{StattTimusk:1999} and organic\cite{Mayaffre:1994} superconductors, namely
the opening of a pairing-fluctuation induced pseudogap.

The structure of this paper is as follows. In section \ref
{Section_Green_Bethe} we introduce the general many-body formalism to obtain
expressions for the self-energy and irreducible vertices in the functional
derivative approach. Although in principle standard, the functional
derivative approach in the particle-particle channel is not widely used. It
allows us to establish a number of exact results. At this level, all the
results could have been obtained with formal diagrammatic expansions in
skeleton diagrams, but we refrain from doing this since the functional
approach is far more economical in this context. The exact results that we
present form the basis of the Two-Particle Self-Consistent (TPSC)\ approach
and its extension (section \ref{Section_TPSC}). These are non-perturbative
approximations that are not diagrammatic resummations. A discussion of the
various exact results that our approach satisfies either exactly or
self-consistently is then presented in section \ref{S_Exact}. Details of the
derivation of the exact results, namely sum rules and high-frequency
expansions, are given in Appendix \ref{Annexe_regle_de_somme}. The last
appendix compares our approach with various other approaches and explains
the connection with the formalism for the repulsive Hubbard model. A summary
of our main results is in the conclusion, section \ref{Section_conclusion}.

\section{Exact relationships between Green function, Self-energy and Vertices%
}

\label{Section_Green_Bethe}

In this section, we derive a number of exact results using the functional
derivative formalism. It is on the basis of these results, and using again
the functional derivative approach, that our approximation scheme will be
developed in section \ref{Section_TPSC}.

\subsection{Definitions, equations of motion}

We work with creation-annihilation operators $\psi _{\uparrow }^{\dagger
},\psi _{\downarrow }$ for Wannier states of spin $\sigma =\uparrow
,\downarrow $ located at position ${\bf r}_{1},$ and, in the Heisenberg
representation, imaginary time $\tau _{1}.$ The space and imaginary time
indices are abbreviated by arabic numerals. Furthermore, we use the Nambu
representation, 
\begin{eqnarray}
\Psi ^{\dagger }\left( 1\right) &=&\left( \psi _{\uparrow }^{\dagger }\left(
1\right) ,\psi _{\downarrow }\left( 1\right) \right) \\
\Psi \left( 1\right) &=&\left( 
\begin{array}{c}
\psi _{\uparrow }\left( 1\right) \\ 
\psi _{\downarrow }^{\dagger }\left( 1\right)
\end{array}
\right)
\end{eqnarray}
where the field operators obey the anticommutation relations 
\begin{equation}
\left\{ \Psi _{\alpha }\left( 1\right) ,\Psi _{\beta }^{\dagger }\left(
2\right) \right\} \delta \left( \tau _{1}-\tau _{2}\right) =\delta \left(
1-2\right) \delta _{\alpha ,\beta }~.
\end{equation}
In this notation, the space part of Dirac delta functions are Kroenecker
deltas: $\delta \left( 1-2\right) =\delta _{{\bf r}_{1},{\bf r}_{2}}\delta
\left( \tau _{1}-\tau _{2}\right) .$ When numerals are set in bold face, we
mean just the space position (e.g. $t\left( {\bf 1,2}\right) $) and refer to
the Schr\"{o}dinger picture. Adding the convention that indices with an
overbar are summed over space positions and, when appropriate, integrated
over imaginary time from $0$ to $\beta ,$ the Hubbard Hamiltonian takes the
form 
\begin{equation}
\widehat{H}=-t\left( \overline{{\bf 1}}{\bf ,}\overline{{\bf 2}}\right)
\left( \psi _{\overline{\sigma }}^{\dagger }\left( \overline{{\bf 1}}\right)
\psi _{\overline{\sigma }}\left( \overline{{\bf 2}}\right) +\psi _{\overline{%
\sigma }}^{\dagger }\left( \overline{{\bf 2}}\right) \psi _{\overline{\sigma 
}}\left( \overline{{\bf 1}}\right) \right) +U\psi _{\uparrow }^{\dagger
}\left( \overline{{\bf 1}}\right) \psi _{\uparrow }\left( \overline{{\bf 1}}%
\right) \psi _{\downarrow }^{\dagger }\left( \overline{{\bf 1}}\right) \psi
_{\downarrow }\left( \overline{{\bf 1}}\right)  \label{Hamiltonien}
\end{equation}
with $t\left( \overline{{\bf 1}}{\bf ,}\overline{{\bf 2}}\right) $ the
hopping matrix elements.

Since we will work in the grand-canonical ensemble, it is convenient to take 
$\widehat{H}-\mu \widehat{N}$ as the time evolution operator in the
Heisenberg representation, with $\mu $ the chemical potential and $\widehat{N%
}$ the number operator. The corresponding equations of motion for the field
operators then are, 
\begin{equation}
\left[ \left( \frac{\partial }{\partial \tau _{1}}-\mu \right) \delta _{{\bf %
r}_{1},{\bf r}\overline{_{2}}}-t\left( {\bf 1,}\overline{{\bf 2}}\right) %
\right] \delta \left( \tau _{1}-\tau _{\overline{2}}\right) \psi _{\sigma
}\left( \overline{2}\right) =-U\psi _{-\sigma }^{\dagger }\left( 1\right)
\psi _{-\sigma }\left( 1\right) \psi _{\sigma }\left( 1\right)
\label{EqMvt_c}
\end{equation}
\begin{equation}
\left[ \left( \frac{\partial }{\partial \tau _{1}}+\mu \right) \delta _{{\bf %
r}_{1},{\bf r}\overline{_{2}}}+t\left( {\bf 1,}\overline{{\bf 2}}\right) %
\right] \delta \left( \tau _{1}-\tau _{\overline{2}}\right) \psi _{\sigma
}^{\dagger }\left( \overline{2}\right) =U\psi _{-\sigma }^{\dagger }\left(
1\right) \psi _{-\sigma }\left( 1\right) \psi _{\sigma }^{\dagger }\left(
1\right) ~.  \label{EqMvt_c+}
\end{equation}

We will also need the momentum space representation 
\begin{equation}
\psi _{\sigma }\left( {\bf 1}\right) =\frac{1}{\sqrt{N}}\sum_{{\bf k}}e^{i%
{\bf k\cdot r}_{1}}c_{{\bf k,\sigma }}~~;~~\psi _{\sigma }^{\dagger }\left( 
{\bf 1}\right) =\frac{1}{\sqrt{N}}\sum_{{\bf k}}e^{-i{\bf k\cdot r}_{1}}c_{%
{\bf k,\sigma }}^{\dagger }
\end{equation}
and the pair operators 
\begin{equation}
\Delta \left( {\bf 1}\right) =\psi _{\uparrow }\left( {\bf 1}\right) \psi
_{\downarrow }\left( {\bf 1}\right) ~~;~~\Delta ^{\dagger }\left( {\bf 1}%
\right) =\psi _{\downarrow }^{\dagger }\left( {\bf 1}\right) \psi _{\uparrow
}^{\dagger }\left( {\bf 1}\right)  \label{Def_Delta}
\end{equation}
with Fourier transform 
\begin{equation}
\Delta _{{\bf q}}=\frac{1}{\sqrt{N}}\sum_{{\bf k}}c_{{\bf k,}\uparrow }c_{-%
{\bf k+q,}\downarrow }~~;~~\Delta _{{\bf q}}^{\dagger }=\frac{1}{\sqrt{N}}%
\sum_{{\bf k}}c_{-{\bf k+q,}\downarrow }^{\dagger }c_{{\bf k,}\uparrow
}^{\dagger }~.
\end{equation}
These pair operators do not include the interaction potential in their
definition. The pair operators obey the equations of motion, 
\begin{equation}
\frac{\partial \Delta _{{\bf q}}}{\partial \tau }=\frac{-1}{\sqrt{N}}\sum_{%
{\bf k}}\left[ \varepsilon _{{\bf k}}+\varepsilon _{-{\bf k+q}}-2\left( \mu -%
\frac{U}{2}\right) \right] c_{{\bf k,}\uparrow }c_{-{\bf k+q,}\downarrow }
\label{EqMvtDelta}
\end{equation}
\begin{equation}
\frac{\partial \Delta _{{\bf q}}^{\dagger }}{\partial \tau }=\frac{1}{\sqrt{N%
}}\sum_{{\bf k}}\left[ \varepsilon _{{\bf k}}+\varepsilon _{-{\bf k+q}%
}-2\left( \mu -\frac{U}{2}\right) \right] c_{-{\bf k+q,}\downarrow
}^{\dagger }c_{{\bf k,}\uparrow }^{\dagger }  \label{EqMvtDelta+}
\end{equation}
where the band dispersion $\varepsilon _{{\bf k}}$ is the Fourier transform
of the hopping matrix elements 
\begin{equation}
-t\left( {\bf 1,2}\right) =\frac{1}{N}\sum_{{\bf k}}e^{i{\bf k\cdot }\left( 
{\bf r}_{1}-{\bf r}_{2}\right) }\varepsilon _{{\bf k}}~.
\end{equation}

\subsection{Green function and self-energy}

As in Ref.\cite{KadanoffBaym:1962}, we work in the grand canonical ensemble
in the presence of auxiliary source fields that are useful in intermediate
steps of the calculations. The source fields are set to zero at the end.
More specifically, we define the expectation value of a general time-ordered
operator $O$ by 
\begin{equation}
\left\langle T_{\tau }\left[ O\right] \right\rangle _{\Theta }\equiv
Z^{-1}\left( \left\{ {\bf \Theta }\right\} \right) Tr\left\{ e^{-\beta
\left( \widehat{H}-\mu \widehat{N}\right) }T_{\tau }\left[ e^{-\Psi
^{\dagger }\left( \overline{1}\right) {\bf \Theta }\left( \overline{1},%
\overline{2}\right) \Psi \left( \overline{2}\right) }O\right] \right\}
\label{DefMoyenne}
\end{equation}
with 
\begin{equation}
Z\left( \left\{ {\bf \Theta }\right\} \right) \equiv Tr\left\{ e^{-\beta
\left( \widehat{H}-\mu \widehat{N}\right) }T_{\tau }\left[ e^{-\Psi
^{\dagger }\left( \overline{1}\right) {\bf \Theta }\left( \overline{1},%
\overline{2}\right) \Psi \left( \overline{2}\right) }\right] \right\}
\label{DefZ}
\end{equation}
where $T_{\tau }$ is the time-ordering operator while the matrix source
field 
\begin{equation}
\Theta \left( 1,2\right) {\bf =}\left( 
\begin{array}{cc}
0 & \theta \left( 1,2\right) \\ 
\theta ^{\ast }\left( 1,2\right) & 0
\end{array}
\right)
\end{equation}
physically corresponds to Cooper pair sources, 
\begin{equation}
\Psi ^{\dagger }\left( \overline{1}\right) {\bf \Theta }\left( \overline{1},%
\overline{2}\right) \Psi \left( \overline{2}\right) =\theta \left( \overline{%
1},\overline{2}\right) \psi _{\uparrow }^{\dagger }\left( \overline{1}%
\right) \psi _{\downarrow }^{\dagger }\left( \overline{2}\right) +\theta
^{\ast }\left( \overline{1},\overline{2}\right) \psi _{\downarrow }\left( 
\overline{1}\right) \psi _{\uparrow }\left( \overline{2}\right) ~.
\end{equation}
The Nambu Green's function is then a functional of the auxiliary field ${\bf %
\Theta }$ defined by 
\begin{equation}
{\bf G}\left( 1,2;\left\{ {\bf \Theta }\right\} \right) \equiv -\left\langle
T_{\tau }\left[ \Psi \left( 1\right) \Psi ^{\dagger }\left( 2\right) \right]
\right\rangle _{{\bf \Theta }}=-\left( 
\begin{array}{cc}
\left\langle T_{\tau }\left[ \psi _{\uparrow }\left( 1\right) \psi
_{\uparrow }^{\dagger }\left( 2\right) \right] \right\rangle _{{\bf \Theta }}
& \left\langle T_{\tau }\left[ \psi _{\uparrow }\left( 1\right) \psi
_{\downarrow }\left( 2\right) \right] \right\rangle _{{\bf \Theta }} \\ 
\left\langle T_{\tau }\left[ \psi _{\downarrow }^{\dagger }\left( 1\right)
\psi _{\uparrow }^{\dagger }\left( 2\right) \right] \right\rangle _{{\bf %
\Theta }} & \left\langle T_{\tau }\left[ \psi _{\downarrow }^{\dagger
}\left( 1\right) \psi _{\downarrow }\left( 2\right) \right] \right\rangle _{%
{\bf \Theta }}
\end{array}
\right)  \label{DefGreen}
\end{equation}
The quantities that will appear at the end of the calculation will, in
analogy with the Green function, be defined in zero source field, namely 
\begin{equation}
{\bf G}\left( 1,2\right) \equiv {\bf G}\left( 1,2;{\bf 0}\right)
=-\left\langle T_{\tau }\left[ \Psi \left( 1\right) \Psi ^{\dagger }\left(
2\right) \right] \right\rangle ~.
\end{equation}

Using the equations of motion (\ref{EqMvt_c}) and following the algebra of
Ref. \cite{KB43}, one finds that 
\begin{equation}
{\bf G}^{-1}\left( 1,2;\left\{ {\bf \Theta }\right\} \right) ={\bf G}%
_{0}^{-1}\left( 1,2\right) -{\bf \Sigma }\left( 1,2;\left\{ {\bf \Theta }%
\right\} \right) -{\bf \Theta }\left( 1,2\right)  \label{G-1}
\end{equation}
where in Matsubara frequency $\left( k_{n}=\left( 2n+1\right) \pi T\right) $
and Fourier space $\left( {\bf k}\right) $, the non-interacting Green
function takes its usual form, 
\begin{equation}
{\bf G}_{0}\left( {\bf k,}ik_{n}\right) =\left( 
\begin{array}{cc}
\frac{1}{ik_{n}-\left( \varepsilon _{{\bf k}}-\mu \right) } & 0 \\ 
0 & \frac{-1}{-ik_{n}-\left( \varepsilon _{-{\bf k}}-\mu \right) }
\end{array}
\right)
\end{equation}
while the effects of interactions are contained in the self-energy matrix
defined by 
\begin{equation}
{\bf \Sigma }\left( 1,\overline{3};\left\{ {\bf \Theta }\right\} \right) 
{\bf G}\left( \overline{3},2;\left\{ {\bf \Theta }\right\} \right) =U\left( 
\begin{array}{cc}
-\left\langle T_{\tau }\left[ \psi _{\downarrow }^{\dagger }\left(
1^{+}\right) \psi _{\downarrow }\left( 1\right) \psi _{\uparrow }\left(
1\right) \psi _{\uparrow }^{\dagger }\left( 2\right) \right] \right\rangle _{%
{\bf \Theta }} & -\left\langle T_{\tau }\left[ \psi _{\downarrow }^{\dagger
}\left( 1^{+}\right) \psi _{\downarrow }\left( 1\right) \psi _{\uparrow
}\left( 1\right) \psi _{\downarrow }\left( 2\right) \right] \right\rangle _{%
{\bf \Theta }} \\ 
\left\langle T_{\tau }\left[ \psi _{\uparrow }^{\dagger }\left( 1^{+}\right)
\psi _{\uparrow }\left( 1\right) \psi _{\downarrow }^{\dagger }\left(
1\right) \psi _{\uparrow }^{\dagger }\left( 2\right) \right] \right\rangle _{%
{\bf \Theta }} & \left\langle T_{\tau }\left[ \psi _{\uparrow }^{\dagger
}\left( 1^{+}\right) \psi _{\uparrow }\left( 1\right) \psi _{\downarrow
}^{\dagger }\left( 1\right) \psi _{\downarrow }\left( 2\right) \right]
\right\rangle _{{\bf \Theta }}
\end{array}
\right) ~.  \label{SelfGeneral}
\end{equation}
The notation $1^{+}$ indicates that the imaginary time is infinitesimally
larger than $\tau _{1},$ $\left( \text{or smaller for }1^{-}\right) .$

\subsection{Self-energy and pair susceptibility}

The self-energy, as should be clear from the last equation, depends on
four-point functions that may be calculated in different channels. For the
repulsive Hubbard model, charge and spin fluctuation channels are dominant,
so approximations for the four point functions are written down in these
channels. However, in the nearest-neighbor attractive model away from
half-filling, it is the pair fluctuations that are dominant, even in the
paramagnetic state. As a preliminary remark, we can suggest how the
self-energy will be related to the pair correlation function by making the
key observation that the diagonal terms can be written as functional
derivatives with respect to the $\theta $ field. First, note that when $%
\theta $ is set to zero in Eq.(\ref{SelfGeneral}) it becomes diagonal and $%
{\bf \Sigma }_{11}$ simplifies to 
\begin{equation}
{\bf \Sigma }_{11}\left( 1,2\right) =-U\left\langle T_{\tau }\left[ \psi
_{\downarrow }^{\dagger }\left( 1^{+}\right) \psi _{\downarrow }\left(
1\right) \psi _{\uparrow }\left( 1\right) \psi _{\uparrow }^{\dagger }\left( 
\overline{3}\right) \right] \right\rangle G_{11}^{-1}\left( \overline{3}%
,2\right) .
\end{equation}
The operators located in the middle, $\psi _{\downarrow }\left( 1\right)
\psi _{\uparrow }\left( 1\right) ,$ can be obtained from a functional
derivative with respect to $\theta ^{\ast }\left( 1,1\right) $ before this
field is set to zero. Hence, taking the functional derivative and setting
off-diagonal terms to zero afterwards, one is left with 
\begin{equation}
{\bf \Sigma }_{11}\left( 1,2\right) =-U\left. \frac{\delta G_{21}\left(
1^{+},\overline{3};\left\{ {\bf \Theta }\right\} \right) }{\delta \theta
^{\ast }\left( 1,1\right) }\right| _{{\bf \Theta }=0}G_{11}^{-1}\left( 
\overline{3},2\right)  \label{Self1}
\end{equation}
This is basically what we want. The self-energy will be written in terms of
a response function in the particle-particle channel.

To continue more generally, we step back and define the susceptibility
matrix 
\begin{equation}
{\bf X}\left( 1,2,3,4;\left\{ {\bf \Theta }\right\} \right) =-\frac{\delta }{%
\delta {\bf \Theta }\left( 4,3\right) }{\bf G}\left( 1,2;\left\{ {\bf \Theta 
}\right\} \right)  \label{SuscepPaireDeriveeFonc}
\end{equation}
where $\delta /\delta {\bf \Theta }\left( 3,4\right) $ is a matrix operator
in Nambu space 
\begin{equation}
\frac{\delta }{\delta {\bf \Theta }\left( 3,4\right) }=\left( 
\begin{array}{cc}
0 & \frac{\delta }{\delta {\bf \theta }^{\ast }\left( 3,4\right) } \\ 
\frac{\delta }{\delta {\bf \theta }\left( 3,4\right) } & 0
\end{array}
\right)
\end{equation}
such that 
\begin{equation}
\frac{\delta }{\delta {\bf \Theta }\left( 3,4\right) }{\bf \Theta }\left(
1,2\right) =\delta \left( 1-3\right) \delta \left( 2-4\right) {\bf I}
\end{equation}
where ${\bf I}$ is the identity matrix in Nambu space. Note that the
susceptibility ${\bf X}$ is still a matrix in Nambu space with only two
matrix indices (four matrix elements). With this notation, the two-point
function that we will need to compute the self-energy from Eq.(\ref
{SelfGeneral}) is the following special case of Eq.(\ref
{SuscepPaireDeriveeFonc}): ${\bf X}\left( 1,2,1^{\mp },1^{\mp };\left\{ {\bf %
\Theta }\right\} \right) .$ Evaluating the functional differentiation
explicitly, we have 
\begin{eqnarray}
{\bf X}\left( 1,2,1^{\mp },1^{\mp };\left\{ {\bf \Theta }\right\} \right)
&\equiv &-\left( 
\begin{array}{cc}
0 & \frac{\delta }{\delta \theta ^{\ast }\left( 1^{-},1^{-}\right) } \\ 
\frac{\delta }{\delta \theta \left( 1^{+},1^{+}\right) } & 0
\end{array}
\right) \left( 
\begin{array}{cc}
-\left\langle T_{\tau }\left[ \psi _{\uparrow }\left( 1\right) \psi
_{\uparrow }^{\dagger }\left( 2\right) \right] \right\rangle _{{\bf \Theta }}
& -\left\langle T_{\tau }\left[ \psi _{\uparrow }\left( 1\right) \psi
_{\downarrow }\left( 2\right) \right] \right\rangle _{{\bf \Theta }} \\ 
-\left\langle T_{\tau }\left[ \psi _{\downarrow }^{\dagger }\left( 1\right)
\psi _{\uparrow }^{\dagger }\left( 2\right) \right] \right\rangle _{{\bf %
\Theta }} & -\left\langle T_{\tau }\left[ \psi _{\downarrow }^{\dagger
}\left( 1\right) \psi _{\downarrow }\left( 2\right) \right] \right\rangle _{%
{\bf \Theta }}
\end{array}
\right) \\
&=&\left( 
\begin{array}{cc}
\frac{\delta }{\delta \theta ^{\ast }\left( 1^{-},1^{-}\right) }\left\langle
T_{\tau }\left[ \psi _{\downarrow }^{\dagger }\left( 1\right) \psi
_{\uparrow }^{\dagger }\left( 2\right) \right] \right\rangle _{{\bf \Theta }}
& \frac{\delta }{\delta \theta ^{\ast }\left( 1^{-},1^{-}\right) }%
\left\langle T_{\tau }\left[ \psi _{\downarrow }^{\dagger }\left( 1\right)
\psi _{\downarrow }\left( 2\right) \right] \right\rangle _{{\bf \Theta }} \\ 
\frac{\delta }{\delta \theta \left( 1^{+},1^{+}\right) }\left\langle T_{\tau
}\left[ \psi _{\uparrow }\left( 1\right) \psi _{\uparrow }^{\dagger }\left(
2\right) \right] \right\rangle _{{\bf \Theta }} & \frac{\delta }{\delta
\theta \left( 1^{+},1^{+}\right) }\left\langle T_{\tau }\left[ \psi
_{\uparrow }\left( 1\right) \psi _{\downarrow }\left( 2\right) \right]
\right\rangle _{{\bf \Theta }}
\end{array}
\right) \\
&=&-\left( 
\begin{array}{c}
\left\langle T_{\tau }\left[ \psi _{\downarrow }^{\dagger }\left( 1\right)
\psi _{\downarrow }\left( 1^{-}\right) \psi _{\uparrow }\left( 1^{-}\right)
\psi _{\uparrow }^{\dagger }\left( 2\right) \right] \right\rangle _{{\bf %
\Theta }} \\ 
\left\langle T_{\tau }\left[ \psi _{\uparrow }\left( 1\right) \psi
_{\uparrow }^{\dagger }\left( 1^{+}\right) \psi _{\downarrow }^{\dagger
}\left( 1^{+}\right) \psi _{\uparrow }^{\dagger }\left( 2\right) \right]
\right\rangle _{{\bf \Theta }}
\end{array}
\begin{array}{c}
\left\langle T_{\tau }\left[ \psi _{\downarrow }^{\dagger }\left( 1\right)
\psi _{\downarrow }\left( 1^{-}\right) \psi _{\uparrow }\left( 1^{-}\right)
\psi _{\downarrow }\left( 2\right) \right] \right\rangle _{{\bf \Theta }} \\ 
\left\langle T_{\tau }\left[ \psi _{\uparrow }\left( 1\right) \psi
_{\uparrow }^{\dagger }\left( 1^{+}\right) \psi _{\downarrow }^{\dagger
}\left( 1^{+}\right) \psi _{\downarrow }\left( 2\right) \right]
\right\rangle _{{\bf \Theta }}
\end{array}
\right) \\
&&-{\bf F}\left( 1^{\mp },1^{\mp };\left\{ {\bf \Theta }\right\} \right) 
{\bf G}\left( 1,2;\left\{ {\bf \Theta }\right\} \right)
\end{eqnarray}
where we defined a function ${\bf F}$ that contains only the anomalous
pieces of the full Green function 
\begin{equation}
{\bf F}\left( 1^{\mp },1^{\mp };\left\{ {\bf \Theta }\right\} \right) \equiv
\left( 
\begin{array}{cc}
0 & -\left\langle T_{\tau }\left[ \psi _{\uparrow }\left( 1^{-}\right) \psi
_{\downarrow }\left( 1^{-}\right) \right] \right\rangle _{{\bf \Theta }} \\ 
-\left\langle T_{\tau }\left[ \psi _{\downarrow }^{\dagger }\left(
1^{+}\right) \psi _{\uparrow }^{\dagger }\left( 1^{+}\right) \right]
\right\rangle _{{\bf \Theta }} & 0
\end{array}
\right) ~.
\end{equation}
The last term in the above expression for the pair susceptibility comes from
functional differentiation of the denominator in the definition of averages,
as one can check with the definition Eq.(\ref{DefZ}) for the partition
function, 
\begin{equation}
\left( 
\begin{array}{cc}
0 & \frac{\delta }{\delta \theta ^{\ast }\left( 1^{-},1^{-}\right) } \\ 
\frac{\delta }{\delta \theta \left( 1^{+},1^{+}\right) } & 0
\end{array}
\right) \frac{1}{Z\left( \left\{ {\bf \Theta }\right\} \right) }=\left( 
\begin{array}{cc}
0 & -\left\langle T_{\tau }\left[ \psi _{\downarrow }\left( 1^{-}\right)
\psi _{\uparrow }\left( 1^{-}\right) \right] \right\rangle _{{\bf \Theta }}
\\ 
-\left\langle T_{\tau }\left[ \psi _{\uparrow }^{\dagger }\left(
1^{+}\right) \psi _{\downarrow }^{\dagger }\left( 1^{+}\right) \right]
\right\rangle _{{\bf \Theta }} & 0
\end{array}
\right) \frac{-1}{Z\left( \left\{ {\bf \Theta }\right\} \right) }~.
\end{equation}

Using rotational invariance and our general result for the self-energy Eq.(%
\ref{SelfGeneral}), we finally find the following key relation between
self-energy and susceptibility 
\begin{equation}
{\bf \Sigma }\left( 1,\overline{3};\left\{ {\bf \Theta }\right\} \right) 
{\bf G}\left( \overline{3},2;\left\{ {\bf \Theta }\right\} \right) =U{\bf X}%
\left( 1,2,1^{\mp },1^{\mp };\left\{ {\bf \Theta }\right\} \right) +U{\bf F}%
\left( 1^{\mp },1^{\mp };\left\{ {\bf \Theta }\right\} \right) {\bf G}\left(
1,2;\left\{ {\bf \Theta }\right\} \right)  \label{SigmaG_External}
\end{equation}
or 
\begin{equation}
{\bf \Sigma }\left( 1,2;\left\{ {\bf \Theta }\right\} \right) =U{\bf X}%
\left( 1,\overline{3},1^{\mp },1^{\mp };\left\{ {\bf \Theta }\right\}
\right) {\bf G}^{-1}\left( \overline{3},2;\left\{ {\bf \Theta }\right\}
\right) +U{\bf F}\left( 1^{\mp },1^{\mp };\left\{ {\bf \Theta }\right\}
\right) \delta \left( 1-2\right) {\bf I~.}  \label{Sigma_External}
\end{equation}

\subsection{Bethe-Salpeter equation for the three-point susceptibility and
relationship between irreducible vertices and self-energy}

In this section, we derive the Bethe-Salpeter equation for the
susceptibility using the functional derivative scheme\cite{KadanoffBaym:1962}%
. This equation allows to define the two-particle irreducible vertex that
plays for the susceptibility a role analogous to that of the self-energy for
the Green function. In deriving the Bethe-Salpeter equation, we will recover
the well known relation between self-energy and particle-particle
irreducible vertices. Since the susceptibility is the functional derivative
of ${\bf G}$ while the self-energy is trivially related to ${\bf G}^{-1},$
it is natural to start from 
\begin{equation}
{\bf G}\left( 1,\overline{4};\left\{ {\bf \Theta }\right\} \right) {\bf G}%
^{-1}\left( \overline{4},5;\left\{ {\bf \Theta }\right\} \right) =\delta
\left( 1-5\right) {\bf I}
\end{equation}
and to take the functional derivative of this equation. One finds, 
\begin{equation}
\frac{\delta }{\delta {\bf \Theta }\left( 2,2\right) }\left[ {\bf G}\left( 1,%
\overline{4};\left\{ {\bf \Theta }\right\} \right) {\bf G}^{-1}\left( 
\overline{4},5;\left\{ {\bf \Theta }\right\} \right) \right] =0
\end{equation}
\begin{eqnarray}
\left[ \frac{\delta }{\delta {\bf \Theta }\left( 2,2\right) }{\bf G}\left( 1,%
\overline{4};\left\{ {\bf \Theta }\right\} \right) \right] {\bf G}%
^{-1}\left( \overline{4},5;\left\{ {\bf \Theta }\right\} \right) &=&-{\bf %
\sigma }_{+}{\bf G}\left( 1,\overline{4}\right) \left( \frac{\delta }{\delta
\theta ^{\ast }\left( 2,2\right) }{\bf G}^{-1}\left( \overline{4},5;\left\{ 
{\bf \Theta }\right\} \right) \right)  \label{dGG-1} \\
&&-{\bf \sigma }_{-}{\bf G}\left( 1,\overline{4}\right) \left( \frac{\delta 
}{\delta \theta \left( 2,2\right) }{\bf G}^{-1}\left( \overline{4},5;\left\{ 
{\bf \Theta }\right\} \right) \right)
\end{eqnarray}
where we used the definitions 
\[
{\bf \sigma }_{+}\equiv \left( 
\begin{array}{cc}
0 & 1 \\ 
0 & 0
\end{array}
\right) \quad ;\quad {\bf \sigma }_{-}\equiv \left( 
\begin{array}{cc}
0 & 0 \\ 
1 & 0
\end{array}
\right) ~. 
\]
Using the expression for the inverse Green function Eq.(\ref{G-1}) and the
chain rule to take into account the dependence of ${\bf \Sigma }$ on $%
\left\{ {\bf \Theta }\right\} $ through\cite{NoteSigmaDeGreen} ${\bf G,}$
one obtains, 
\begin{equation}
\frac{\delta }{\delta \theta ^{\ast }\left( 2,2\right) }{\bf G}^{-1}\left(
4,5;\left\{ {\bf \Theta }\right\} \right) =-\delta \left( 2-5\right) \delta
\left( 2-4\right) {\bf \sigma }_{-}-\frac{\delta {\bf \Sigma }\left(
4,5;\left\{ {\bf \Theta }\right\} \right) }{\delta G_{\overline{k}\overline{%
\ell }}\left( \overline{6},\overline{7};\left\{ {\bf \Theta }\right\}
\right) }\frac{\delta G_{\overline{k}\overline{\ell }}\left( \overline{6},%
\overline{7};\left\{ {\bf \Theta }\right\} \right) }{\delta \theta ^{\ast
}\left( 2,2\right) }  \label{BS_*}
\end{equation}
\begin{equation}
\frac{\delta }{\delta \theta \left( 2,2\right) }{\bf G}^{-1}\left(
4,5;\left\{ {\bf \Theta }\right\} \right) =-\delta \left( 2-5\right) \delta
\left( 2-4\right) {\bf \sigma }_{+}-\frac{\delta {\bf \Sigma }\left(
4,5;\left\{ {\bf \Theta }\right\} \right) }{\delta G_{\overline{k}\overline{%
\ell }}\left( \overline{6},\overline{7};\left\{ {\bf \Theta }\right\}
\right) }\frac{\delta G_{\overline{k}\overline{\ell }}\left( \overline{6},%
\overline{7};\left\{ {\bf \Theta }\right\} \right) }{\delta \theta \left(
2,2\right) }~.  \label{BS_}
\end{equation}

The above equations simplify greatly when the functional derivative is
evaluated in the normal, zero source field case, where particle number
conservation implies the vanishing of anomalous correlation functions. We
are left with, 
\begin{equation}
\left. \frac{\delta G_{\overline{k}\overline{\ell }}\left( 6,7;\left\{ {\bf %
\Theta }\right\} \right) }{\delta \theta ^{\ast }\left( 2,2\right) }\right|
_{{\bf \Theta =0}}=\delta _{\overline{k},2}\delta _{\overline{\ell }%
,1}\left. \frac{\delta G_{21}\left( 6,7;\left\{ {\bf \Theta }\right\}
\right) }{\delta \theta ^{\ast }\left( 2,2\right) }\right| _{{\bf \Theta =0}}
\end{equation}
\begin{equation}
\left. \frac{\delta G_{\overline{k}\overline{\ell }}\left( 6,7;\left\{ {\bf %
\Theta }\right\} \right) }{\delta \theta \left( 2,2\right) }\right| _{{\bf %
\Theta =0}}=\delta _{\overline{k},1}\delta _{\overline{\ell },2}\left. \frac{%
\delta G_{12}\left( 6,7;\left\{ {\bf \Theta }\right\} \right) }{\delta
\theta \left( 2,2\right) }\right| _{{\bf \Theta =0}}~.
\end{equation}
Using the latter results in Eq.(\ref{dGG-1}) multiplying it by ${\bf G}%
\left( 5,3\right) $ and then integrating over the point $5$, we find, with
the help of Eqs.(\ref{BS_*}) and (\ref{BS_}), 
\begin{eqnarray}
\left. \frac{\delta }{\delta {\bf \Theta }\left( 2,2\right) }{\bf G}\left(
1,3;\left\{ {\bf \Theta }\right\} \right) \right| _{{\bf \Theta =0}} &=&{\bf %
\sigma }_{+}{\bf G}\left( 1,2\right) {\bf \sigma }_{-}{\bf G}\left(
2,3\right) +{\bf \sigma }_{-}{\bf G}\left( 1,2\right) {\bf \sigma }_{+}{\bf G%
}\left( 2,3\right)  \label{Bethe-SalpeterGen} \\
&&+{\bf \sigma }_{+}{\bf G}\left( 1,\overline{4}\right) \left. \frac{\delta 
{\bf \Sigma }\left( \overline{4},\overline{5};\left\{ {\bf \Theta }\right\}
\right) }{\delta G_{21}\left( \overline{6},\overline{7}\right) }\right| _{%
{\bf \Theta =0}}\left. \frac{\delta G_{21}\left( \overline{6},\overline{7}%
;\left\{ {\bf \Theta }\right\} \right) }{\delta \theta ^{\ast }\left(
2,2\right) }\right| _{{\bf \Theta =0}}{\bf G}\left( \overline{5},3\right) \\
&&+{\bf \sigma }_{-}{\bf G}\left( 1,\overline{4}\right) \left. \frac{\delta 
{\bf \Sigma }\left( \overline{4},\overline{5};\left\{ {\bf \Theta }\right\}
\right) }{\delta G_{12}\left( \overline{6},\overline{7}\right) }\right| _{%
{\bf \Theta =0}}\left. \frac{\delta G_{12}\left( \overline{6},\overline{7}%
;\left\{ {\bf \Theta }\right\} \right) }{\delta \theta \left( 2,2\right) }%
\right| _{{\bf \Theta =0}}{\bf G}\left( \overline{5},3\right) ~.
\end{eqnarray}
Since we consider the normal phase, the off-diagonal Green functions vanish
and we are left with only two equations that come from the diagonal
components of the above matrix equation. The off-diagonal parts just tell us
that two of the irreducible vertices vanish. Finally then, 
\begin{equation}
\left. \frac{\delta G_{21}\left( 1,3;\left\{ {\bf \Theta }\right\} \right) }{%
\delta \theta ^{\ast }\left( 2,2\right) }\right| _{{\bf \Theta =0}%
}=G_{22}\left( 1,2\right) G_{11}\left( 2,3\right) +G_{22}\left( 1,\overline{4%
}\right) \left. \frac{\delta \Sigma _{21}\left( \overline{4},\overline{5}%
;\left\{ {\bf \Theta }\right\} \right) }{\delta G_{21}\left( \overline{6},%
\overline{7}\right) }\right| _{{\bf \Theta =0}}\left. \frac{\delta
G_{21}\left( \overline{6},\overline{7};\left\{ {\bf \Theta }\right\} \right) 
}{\delta \theta ^{\ast }\left( 2,2\right) }\right| _{{\bf \Theta =0}%
}G_{11}\left( \overline{5},3\right)  \label{BetheSalTheta*}
\end{equation}
\begin{equation}
\left. \frac{\delta G_{12}\left( 1,3;\left\{ {\bf \Theta }\right\} \right) }{%
\delta \theta \left( 2,2\right) }\right| _{{\bf \Theta =0}}=G_{11}\left(
1,2\right) G_{22}\left( 2,3\right) +G_{11}\left( 1,\overline{4}\right)
\left. \frac{\delta \Sigma _{12}\left( \overline{4},\overline{5};\left\{ 
{\bf \Theta }\right\} \right) }{\delta G_{12}\left( \overline{6},\overline{7}%
\right) }\right| _{{\bf \Theta =0}}\left. \frac{\delta G_{12}\left( 
\overline{6},\overline{7};\left\{ {\bf \Theta }\right\} \right) }{\delta
\theta \left( 2,2\right) }\right| _{{\bf \Theta =0}}G_{22}\left( \overline{5}%
,3\right) ~.  \label{BetheSalTheta}
\end{equation}

\subsection{An exact expression for the self-energy where low- and
high-frequency behaviors are separated.}

\label{SSHauteFrequence}

The high-frequency limit of the self-energy is given by the Hartree-Fock
result, as can be shown from sum rules. For latter purposes in our
approximative scheme, it is useful to have at hand an exact expression for
the self-energy where the high-frequency behavior appears explicitly. In
this section, we derive such an expression in the case where the auxiliary
field ${\bf \Theta }$ vanishes.

First, let us recall the sum rules that fix the high-frequency behavior of
the self-energy. In the absence of external field 
\begin{equation}
{\bf G}\left( {\bf k,}ik_{n}\right) =\int \frac{d\omega }{2\pi }\frac{{\bf A}%
\left( {\bf k,}\omega \right) }{ik_{n}-\omega }
\end{equation}
where the single-particle spectral weight ${\bf A}\left( {\bf k,}\omega
\right) $ for the Nambu Green function is given by, 
\begin{equation}
{\bf A}\left( {\bf k,}\omega \right) =\int dte^{i\omega t}\left( 
\begin{array}{cc}
\left\langle \left\{ c_{{\bf k,\uparrow }}\left( t\right) ,c_{{\bf %
k,\uparrow }}^{\dagger }\right\} \right\rangle & 0 \\ 
0 & \left\langle \left\{ c_{{\bf k,\downarrow }}^{\dagger }\left( t\right)
,c_{{\bf k,\downarrow }}\right\} \right\rangle
\end{array}
\right) ~.
\end{equation}
The high-frequency expansion of the Green function is then 
\begin{equation}
{\bf G}\left( {\bf k,}ik_{n}\right) =\frac{1}{ik_{n}}\int \frac{d\omega }{%
2\pi }{\bf A}\left( {\bf k,}\omega \right) +\frac{1}{\left( ik_{n}\right)
^{2}}\int \frac{d\omega }{2\pi }\omega {\bf A}\left( {\bf k,}\omega \right)
+\ldots  \label{G(k,w)_haute_frequence}
\end{equation}
where the frequency moments of ${\bf A}\left( {\bf k,}\omega \right) $ are
easily computed from equal-time commutators 
\begin{equation}
\int \frac{d\omega }{2\pi }{\bf A}\left( {\bf k,}\omega \right) ={\bf I}
\label{A(k,w)_normalisation}
\end{equation}
\begin{equation}
\int \frac{d\omega }{2\pi }\omega {\bf A}\left( {\bf k,}\omega \right)
=\left( \varepsilon _{{\bf k}}-\mu \right) \sigma _{z}+U\left( 
\begin{array}{cc}
\left\langle n_{\downarrow }\right\rangle & 0 \\ 
0 & -\left\langle n_{\uparrow }\right\rangle
\end{array}
\right) ~.  \label{A(k,w)_premier_moment}
\end{equation}
Comparing with 
\begin{eqnarray}
{\bf G}\left( {\bf k,}ik_{n}\right) &=&\left( ik_{n}{\bf I-}\left(
\varepsilon _{{\bf k}}-\mu \right) \sigma _{z}-{\bf \Sigma }\left( {\bf k,}%
ik_{n}\right) \right) ^{-1}  \nonumber \\
&=&\frac{1}{ik_{n}}{\bf I}+\frac{1}{\left( ik_{n}\right) ^{2}}\left[ \left(
\varepsilon _{{\bf k}}-\mu \right) \sigma _{z}+{\bf \Sigma }\left( {\bf k,}%
ik_{n}\right) \right] +\ldots
\end{eqnarray}
it is clear, using spin-rotational invariance, that 
\begin{equation}
\lim_{ik_{n}\rightarrow \infty }{\bf \Sigma }\left( {\bf k,}ik_{n}\right)
=U\left\langle n_{\downarrow }\right\rangle \sigma _{z}~.  \label{SigmaHF}
\end{equation}

An expression where this asymptotic behavior is explicit may easily be
obtained from the general formula for the self-energy specialized to zero
external field, namely Eq.(\ref{Sigma_External}) with $\Theta =0.$ Returning
to our general discussion Eq.(\ref{Self1}) will help understand the point we
were making. In zero external field, we have 
\begin{equation}
{\bf \Sigma }\left( 1,2\right) =U{\bf X}\left( 1,\overline{3},1^{\mp
},1^{\mp }\right) {\bf G}^{-1}\left( \overline{3},2\right) =-U\left. \frac{%
\delta }{\delta {\bf \Theta }\left( 1^{\mp },1^{\mp }\right) }{\bf G}\left(
1,\overline{3};\left\{ {\bf \Theta }\right\} \right) \right| _{\Theta =0}%
{\bf G}^{-1}\left( \overline{3},2\right)
\end{equation}
or, looking only at the non-vanishing elements, 
\begin{eqnarray}
\Sigma _{11}\left( 1,2\right) &=&-U\left. \frac{\delta G_{21}\left( 1^{+},%
\overline{3};\left\{ {\bf \Theta }\right\} \right) }{\delta \theta ^{\ast
}\left( 1,1\right) }\right| _{{\bf \Theta }=0}G_{11}^{-1}\left( \overline{3}%
,2\right)  \label{Sig11etTr(SG)} \\
\Sigma _{22}\left( 1,2\right) &=&-U\left. \frac{\delta G_{12}\left( 1,%
\overline{3};\left\{ {\bf \Theta }\right\} \right) }{\delta \theta \left(
1^{+},1^{+}\right) }\right| _{{\bf \Theta }=0}G_{22}^{-1}\left( \overline{3}%
,2\right)
\end{eqnarray}
Substituting the Bethe-Salpeter equations (\ref{BetheSalTheta*})(\ref
{BetheSalTheta}) in the last two expressions we have the equivalent
expressions, 
\begin{equation}
\Sigma _{11}\left( 1,2\right) =-UG_{22}\left( 1^{+},1\right) \delta \left(
1-2\right) -UG_{22}\left( 1^{+},\overline{4}\right) \left. \frac{\delta
\Sigma _{21}\left( \overline{4},2;\left\{ {\bf \Theta }\right\} \right) }{%
\delta G_{21}\left( \overline{6},\overline{7}\right) }\right| _{{\bf \Theta
=0}}\left. \frac{\delta G_{21}\left( \overline{6},\overline{7};\left\{ {\bf %
\Theta }\right\} \right) }{\delta \theta ^{\ast }\left( 1,1\right) }\right|
_{{\bf \Theta =0}}  \label{Self_Exacte_HFexplicite11}
\end{equation}
\begin{equation}
\Sigma _{22}\left( 1,2\right) =-UG_{11}\left( 1,1^{+}\right) \delta \left(
1^{+}-2\right) -UG_{11}\left( 1,\overline{4}\right) \left. \frac{\delta
\Sigma _{12}\left( \overline{4},2;\left\{ {\bf \Theta }\right\} \right) }{%
\delta G_{12}\left( \overline{6},\overline{7}\right) }\right| _{{\bf \Theta
=0}}\left. \frac{\delta G_{12}\left( \overline{6},\overline{7};\left\{ {\bf %
\Theta }\right\} \right) }{\delta \theta \left( 1^{+},1^{+}\right) }\right|
_{{\bf \Theta =0}}~.  \label{Self_Exacte_HFexplicite22}
\end{equation}
One may easily check that the terms proportional to $\delta \left(
1-2\right) $ in these two exact expressions are precisely the Hartree-Fock
contribution, Eq.(\ref{SigmaHF}).

The skeleton diagram representation of the Bethe-Salpeter equation (\ref
{BetheSalTheta*}) and of the self-energy Eq.(\ref{Self_Exacte_HFexplicite11}%
) appear in Fig.1 of the accompanying paper. This skeleton diagram
representation is not necessary to understand the rest of the paper but may
be useful for physical understanding. The non-perturbative approach
developed in the following section does not directly correspond to the
summation of an infinite subset of diagrams.

\section{A systematic non-perturbative approach}

\label{Section_TPSC}

A brief reminder of how Hartree-Fock theory is derived in the functional
derivative approach will motivate the two-particle self-consistent approach.
At this first level of approximation (TPSC), one has ${\bf G}^{\left(
1\right) },$ ${\bf \Sigma }^{\left( 1\right) }$ in the presence of ${\bf %
\Theta }$ and the corresponding irreducible vertices and susceptibilities
are obtained from the functional derivative approach. The only unknown
quantity, double occupancy $\left\langle n_{\uparrow }n_{\downarrow
}\right\rangle $, may be obtained self-consistently by using what we will
call the local-pair sum rule. The local-pair sum rule is a simple
consequence of the fluctuation-dissipation theorem. An improved
approximation for the self-energy, ${\bf \Sigma }^{\left( 2\right) }$, will
be found in section \ref{Sec_TPSC_self(2)}. We conclude with a discussion of
an internal accuracy check that, as in the repulsive case, helps delineate
the domain of validity of the approach.

\subsection{A perspective: Conserving approximations and Hartree-Fock theory}

In conserving approximations, the self-energy is obtained from a functional
derivative of the Luttinger-Ward functional\cite{LuttingerWard:1960} that is
computed from skeleton diagrams. Irreducible vertices for response functions
are then obtained from appropriate functional derivatives. The Hartree-Fock
approach is a special case of conserving approximation. The more standard
way to derive the Hartree-Fock approach is to treat the general equation for
the self-energy Eq.(\ref{SelfGeneral}) in the presence of the auxiliary
field as if Wick's theorem applied to the right-hand side of that equation.
More specifically, Eq.(\ref{SelfGeneral}) becomes, in the Hartree-Fock
approximation 
\begin{equation}
{\bf \Sigma }^{HF}\left( 1,\overline{3};\left\{ {\bf \Theta }\right\}
\right) {\bf G}^{HF}\left( \overline{3},2;\left\{ {\bf \Theta }\right\}
\right) =U\left( 
\begin{array}{cc}
-G_{22}^{HF}\left( 1^{+},1;\left\{ {\bf \Theta }\right\} \right) & 
G_{12}^{HF}\left( 1,1;\left\{ {\bf \Theta }\right\} \right) \\ 
G_{21}^{HF}\left( 1,1;\left\{ {\bf \Theta }\right\} \right) & 
-G_{11}^{HF}\left( 1,1^{+};\left\{ {\bf \Theta }\right\} \right)
\end{array}
\right) {\bf G}^{HF}\left( 1,2;\left\{ {\bf \Theta }\right\} \right)
\label{SigmaG_factoriseHF}
\end{equation}
or 
\begin{equation}
{\bf \Sigma }^{HF}\left( 1,2;\left\{ {\bf \Theta }\right\} \right) =U\left( 
\begin{array}{cc}
-G_{22}^{HF}\left( 1^{+},1;\left\{ {\bf \Theta }\right\} \right) & 
G_{12}^{HF}\left( 1,1;\left\{ {\bf \Theta }\right\} \right) \\ 
G_{21}^{HF}\left( 1,1;\left\{ {\bf \Theta }\right\} \right) & 
-G_{11}^{HF}\left( 1,1^{+};\left\{ {\bf \Theta }\right\} \right)
\end{array}
\right) \delta \left( 1-2\right) ~.
\end{equation}
The corresponding irreducible vertices for the Bethe-Salpeter equation
governing the pair fluctuations, Eq.(\ref{Bethe-SalpeterGen}) are 
\begin{equation}
\left. \frac{\delta \Sigma _{12}^{HF}\left( 1,2;\left\{ {\bf \Theta }%
\right\} \right) }{\delta G_{12}^{HF}\left( 3,4;\left\{ {\bf \Theta }%
\right\} \right) }\right| _{{\bf \Theta =0}}=\left. \frac{\delta \Sigma
_{21}^{HF}\left( 1,2;\left\{ {\bf \Theta }\right\} \right) }{\delta
G_{21}^{HF}\left( 3,4;\left\{ {\bf \Theta }\right\} \right) }\right| _{{\bf %
\Theta =0}}=U\delta \left( 1-2\right) \delta \left( 1-3\right) \delta \left(
1-4\right)
\end{equation}
which leads to the simplest T-matrix type approximation. In the final
calculations, we only need the self-energy in zero field. There it is purely
diagonal and given by 
\begin{equation}
{\bf \Sigma }^{HF}\left( 1,2\right) =\sigma _{z}U\left\langle n_{\downarrow
}\right\rangle \delta \left( 1-2\right) ~.  \label{HartreeFockSimple}
\end{equation}

Returning momentarily to the original equation that was approximated, Eq.(%
\ref{SelfGeneral}), it is important to realize that the Hartree-Fock
approximation satisfies an exact result. First note that the four point
function becomes simple when point $2$ becomes the same as $1.$ In that
case, this quantity takes a simple form that, because of the anticommutation
relations, depends on whether $2\rightarrow 1^{+}$ or $2\rightarrow 1^{-}$.
More specifically, we have the exact results 
\begin{equation}
{\bf \Sigma }\left( 1,\overline{3};\left\{ {\bf \Theta }\right\} \right) 
{\bf G}\left( \overline{3},1^{+};\left\{ {\bf \Theta }\right\} \right)
=U\left( 
\begin{array}{cc}
\left\langle n_{\uparrow }\left( 1\right) n_{\downarrow }\left( 1\right)
\right\rangle _{{\bf \Theta }} & G_{12}\left( 1,1;\left\{ {\bf \Theta }%
\right\} \right) \\ 
0 & \left\langle n_{\uparrow }\left( 1\right) \left( n_{\downarrow }\left(
1\right) -1\right) \right\rangle _{{\bf \Theta }}
\end{array}
\right)  \label{SigmaG+}
\end{equation}
\begin{equation}
{\bf \Sigma }\left( 1,\overline{3};\left\{ {\bf \Theta }\right\} \right) 
{\bf G}\left( \overline{3},1^{-};\left\{ {\bf \Theta }\right\} \right)
=U\left( 
\begin{array}{cc}
\left\langle n_{\downarrow }\left( 1\right) \left( n_{\uparrow }\left(
1\right) -1\right) \right\rangle _{{\bf \Theta }} & 0 \\ 
-G_{21}\left( 1,1;\left\{ {\bf \Theta }\right\} \right) & \left\langle
n_{\uparrow }\left( 1\right) n_{\downarrow }\left( 1\right) \right\rangle _{%
{\bf \Theta }}
\end{array}
\right) ~.  \label{SigmaG-}
\end{equation}
In zero field, the difference of these two exact results is 
\begin{equation}
{\bf \Sigma }\left( 1,\overline{3}\right) {\bf G}\left( \overline{3}%
,1^{+}\right) -{\bf \Sigma }\left( 1,\overline{3}\right) {\bf G}\left( 
\overline{3},1^{-}\right) =\sigma _{z}U\left\langle n_{\downarrow
}\right\rangle ~.  \label{SigmaG_Diff}
\end{equation}
Given what we have said in section \ref{SSHauteFrequence} about the exact
high-frequency behavior of the self-energy, the reader will not be surprised
to learn that the Hartree-Fock approximation does satisfy the relation Eq.(%
\ref{SigmaG_Diff}). Indeed, this exact relation is sensitive only to the
high-frequency limit of the self-energy, as may be proven by writing down
explicitly the $1,1$ component of Eq.(\ref{SigmaG_Diff}) in
Fourier-Matsubara space as follows 
\begin{equation}
+\frac{T}{N}\sum_{{\bf k}}\sum_{ik_{n}}\left( \frac{\Sigma \left( {\bf k}%
,ik_{n}\right) }{ik_{n}-\left( \varepsilon _{{\bf k}}-\mu \right) -\Sigma
\left( {\bf k},ik_{n}\right) }-\frac{U\left\langle n_{\downarrow
}\right\rangle }{ik_{n}}\right) \left(
e^{-ik_{n}0^{-}}-e^{-ik_{n}0^{+}}\right)  \label{SigmaG_et_convergence}
\end{equation}
\begin{equation}
\frac{T}{N}\sum_{{\bf k}}\sum_{ik_{n}}\left[ \frac{U\left\langle
n_{\downarrow }\right\rangle }{ik_{n}}\left(
e^{-ik_{n}0^{-}}-e^{-ik_{n}0^{+}}\right) \right] =U\left\langle
n_{\downarrow }\right\rangle
\end{equation}
In this expression, the first sum vanishes because we have added and
subtracted a term that makes it convergent without the need for convergence
factors $e^{-ik_{n}0^{\pm }}$. Hence, only the last sum survives. The result
is a direct manifestation of the anticommutation relations in the asymptotic
behavior of the Green function.

\subsection{Two-particle Self-Consistency and irreducible vertex}

\label{Two-particle_Self-Consistency}

We will call the two exact results Eqs.(\ref{SigmaG+}) and (\ref{SigmaG-}) Tr%
$\left[ \Sigma G\right] $ for short cut. They are simply related to the
potential energy (and hence to double occupancy), a crucial quantity for the
Hubbard Hamiltonian. Furthermore, they can be considered as initial
conditions for the true expression that defines the self-energy in the most
general case, ${\bf \Sigma }\left( 1,\overline{3};\left\{ {\bf \Theta }%
\right\} \right) {\bf G}\left( \overline{3},2;\left\{ {\bf \Theta }\right\}
\right) $ where point $2$ has moved away from point one. Hence, in the first
step of the approximation that we propose, we perform, as in conserving
approximations, a Hartree-Fock like factorization in an external field, but
we add the constraint that the factorization becomes exact when $%
2\rightarrow 1^{+}$ or $2\rightarrow 1^{-}.$ More specifically, in analogy
with the factorization Eq.(\ref{SigmaG_factoriseHF}) we start from the {\it %
ansatz} 
\begin{equation}
{\bf \Sigma }^{\left( 1\right) }\left( 1,\overline{3};\left\{ {\bf \Theta }%
\right\} \right) {\bf G}^{\left( 1\right) }\left( \overline{3},2;\left\{ 
{\bf \Theta }\right\} \right) =U\left( 
\begin{array}{cc}
-G_{22}^{\left( 1\right) }\left( 1^{+},1;\left\{ {\bf \Theta }\right\}
\right) & G_{12}^{\left( 1\right) }\left( 1,1;\left\{ {\bf \Theta }\right\}
\right) \\ 
G_{21}^{\left( 1\right) }\left( 1,1;\left\{ {\bf \Theta }\right\} \right) & 
-G_{11}^{\left( 1\right) }\left( 1,1^{+};\left\{ {\bf \Theta }\right\}
\right)
\end{array}
\right) {\bf A}\left( \left\{ {\bf \Theta }\right\} \right) {\bf G}^{\left(
1\right) }\left( 1,2;\left\{ {\bf \Theta }\right\} \right)
\label{TPSC_Factorise}
\end{equation}
where the matrix ${\bf A}\left( \left\{ {\bf \Theta }\right\} \right) $ is
obtained by requiring that the exact relations Eqs.(\ref{SigmaG+}) and (\ref
{SigmaG-}) be satisfied\cite{NoteAnsatzAlt}. Setting alternatively $%
2\rightarrow 1^{+}$ or $2\rightarrow 1^{-}$ in the last expression, and
requiring equality with the respective exact Tr$\left[ \Sigma G\right] $
expression, we find, using spin rotational invariance, that in either case
there is a unique solution for the off diagonal elements 
\begin{mathletters}
\begin{equation}
{\bf \Sigma }_{12}^{\left( 1\right) }\left( 1,2;\left\{ {\bf \Theta }%
\right\} \right) =U\frac{\left\langle n_{\uparrow }\left( 1-n_{\downarrow
}\right) \right\rangle _{{\bf \Theta }}G_{12}^{\left( 1\right) }\left(
1,1;\left\{ {\bf \Theta }\right\} \right) \delta \left( 1-2\right) }{%
\left\langle n_{\uparrow }\right\rangle _{{\bf \Theta }}\left\langle
1-n_{\downarrow }\right\rangle _{{\bf \Theta }}-G_{12}^{\left( 1\right)
}\left( 1,1;\left\{ {\bf \Theta }\right\} \right) G_{21}^{\left( 1\right)
}\left( 1,1;\left\{ {\bf \Theta }\right\} \right) }  \label{TPSC_Self}
\end{equation}
with the analogous expression when the Nambu matrix indices $1$ and $2$ are
inverted. It should be clear from this result that the functional dependence
of ${\bf A}\left( \left\{ {\bf \Theta }\right\} \right) $ on external field
is only through ${\bf G}^{\left( 1\right) }\left( 1,1^{\pm };\left\{ {\bf %
\Theta }\right\} \right) $ and double occupancy $\left\langle n_{\downarrow
}n_{\uparrow }\right\rangle _{{\bf \Theta }}.$ Nevertheless, this
establishes a strong self-consistency relation between one and two-particle
quantities that is absent from any standard diagrammatic approach. That is
why we called this portion of the approach ``Two Particle Self-Consistent''
(TPSC)\cite{Vilk:1995}. It is important also to note that the superscript $%
\left( 1\right) $ refers to what, in earlier publications,\cite{Vilk:1997}
we called the zeroth step of the approximation. As we will see in a moment,
at this level of approximation the diagonal part of ${\bf \Sigma }^{\left(
1\right) }$ is a constant when ${\bf \Theta =0}$. Hence, in this limit, $%
{\bf G}^{\left( 1\right) }$ is equal to a bare propagator once the chemical
potential is adjusted to obtain the proper filling. That is why we had
referred to this level of approximation with superscript $\left( 0\right) $
in earlier publications. It is important to notice however that ${\bf \Sigma 
}^{\left( 1\right) }$ (or ${\bf \Sigma }^{\left( 0\right) }$ in the former
notation) may have a dependence on external magnetic field, for example,
that is absent in the non-interacting case.

The particle-particle irreducible vertices appearing in the Bethe-Salpeter
Eqs.(\ref{BetheSalTheta}) and (\ref{BetheSalTheta*}) are obtained from
functional differentiation of the self-energy Eq.(\ref{TPSC_Self}) as in any
conserving approximation. Since when ${\bf \Theta }=0$ all off diagonal
function such as $G_{12}^{\left( 1\right) }\left( 1,1\right) ,G_{21}^{\left(
1\right) }\left( 1,1\right) $ or $\delta \left\langle n_{\downarrow }\left(
1-n_{\uparrow }\right) \right\rangle _{{\bf \Theta }}/\delta G_{12}^{\left(
1\right) }\left( 1,1;\left\{ {\bf \Theta }\right\} \right) $ and the like
vanish, we are left with 
\end{mathletters}
\begin{eqnarray}
\left. \frac{\delta \Sigma _{12}^{\left( 1\right) }\left( 1,2;\left\{ {\bf %
\Theta }\right\} \right) }{\delta G_{12}^{\left( 1\right) }\left(
3,4;\left\{ {\bf \Theta }\right\} \right) }\right| _{{\bf \Theta =0}}
&=&\left. \frac{\delta \Sigma _{21}^{\left( 1\right) }\left( 1,2;\left\{ 
{\bf \Theta }\right\} \right) }{\delta G_{21}^{\left( 1\right) }\left(
3,4;\left\{ {\bf \Theta }\right\} \right) }\right| _{{\bf \Theta =0}}=U\frac{%
\left\langle n_{\downarrow }\left( 1-n_{\uparrow }\right) \right\rangle }{%
\left\langle n_{\downarrow }\right\rangle \left\langle 1-n_{\uparrow
}\right\rangle }\delta \left( 1-2\right) \delta \left( 1-3\right) \delta
\left( 1-4\right)  \label{TPSC_vertex} \\
&\equiv &U_{pp}\delta \left( 1-2\right) \delta \left( 1-3\right) \delta
\left( 1-4\right) ~.
\end{eqnarray}
The irreducible vertex $U_{pp}$ is still local, as in the Hartree-Fock
approximation, but, contrary to Hartree-Fock, the bare vertex is dressed.
The function that dresses the vertex is simply related to double occupancy.
It can be determined from the Bethe-Salpeter equation itself by using the
fluctuation-dissipation theorem, allowing us to close the system of
equations.

\subsection{An approximate expression for $\Sigma ^{\left( 1\right) }$ in
the TPSC approach}

The self-energy entering ${\bf G}^{\left( 1\right) }$ in zero external field
is diagonal. There is however an apparent paradox. Indeed, substituting the
TPSC factorization Eq.(\ref{TPSC_Factorise}) on the left-hand side of the
two Tr$\left[ \Sigma G\right] $ equation Eqs.(\ref{SigmaG+}) and (\ref
{SigmaG-}), all for ${\bf \Theta =0}$, seems to give two different solutions
for the diagonal value of ${\bf A}\left( \left\{ {\bf \Theta }\right\}
\right) ,$ and hence for the corresponding ${\bf \Sigma }^{\left( 1\right) }$%
. Let us use ${\bf \Sigma }^{\left( 1\right) }$ with a $+$ or $-$ index
depending on whether they are the solution of Eqs.(\ref{SigmaG+}) or (\ref
{SigmaG-}). In either case, ${\bf \Sigma }^{\left( 1\right) }\left(
1-2\right) $ is proportional to $\delta \left( 1-2\right) ,$ so that using $%
G_{11\pm }^{\left( 1\right) }\left( 1,1^{+}\right) =\left\langle n_{\uparrow
}\left( 1\right) \right\rangle $ and $G_{11\pm }^{\left( 1\right) }\left(
1,1^{-}\right) =-1+\left\langle n_{\uparrow }\left( 1\right) \right\rangle $
and the like for $G_{22}^{\left( 1\right) },$ one finds, 
\begin{equation}
\Sigma _{11+}^{\left( 1\right) }\left( 1,2\right) =-\Sigma _{22-}^{\left(
1\right) }\left( 1,2\right) =\left( U-U_{pp}\left\langle 1-n_{\downarrow
}\right\rangle \right) \delta \left( 1-2\right)
\end{equation}
\begin{equation}
\Sigma _{11-}^{\left( 1\right) }\left( 1,2\right) =-\Sigma _{22+}^{\left(
1\right) }\left( 1,2\right) =U_{pp}\left\langle n_{\uparrow }\right\rangle
\delta \left( 1-2\right)
\end{equation}
From these results, keeping the filling fixed we have 
\begin{equation}
{\bf \Sigma }_{+}^{\left( 1\right) }\left( 1,\overline{3}\right)
G_{+}^{\left( 1\right) }\left( \overline{3},1^{+}\right) -{\bf \Sigma }%
_{-}^{\left( 1\right) }\left( 1,\overline{3}\right) {\bf G}_{-}^{\left(
1\right) }\left( \overline{3},1^{-}\right) =\sigma _{z}U\left\langle
n_{\uparrow }\left( 1\right) \right\rangle
\end{equation}
This suggests that the antisymmetric combination ${\bf \Sigma }_{+}^{\left(
1\right) }-{\bf \Sigma }_{+}^{\left( 1\right) }$ is related to the
high-frequency asymptotic behavior of the self-energy, as in the
Hartree-Fock case. The symmetric combination on the other hand should not
depend on convergence factors, as can be seen by using arguments analogous
to those of Eq.(\ref{SigmaG_et_convergence}). This quantity should then be a
measure of the low-frequency behavior of the self-energy. With $n\equiv
\left\langle n_{\downarrow }\right\rangle +\left\langle n_{\uparrow
}\right\rangle $ the quantity 
\begin{equation}
\frac{1}{2}\left( {\bf \Sigma }_{+}^{\left( 1\right) }\left( 1,2\right) +%
{\bf \Sigma }_{-}^{\left( 1\right) }\left( 1,2\right) \right) =\sigma
_{z}\left( \frac{U}{2}-\frac{U_{pp}\left( 1-n\right) }{2}\right) \delta
\left( 1-2\right)  \label{Self_ordre_0_et_pot_chim}
\end{equation}
we expect, plays the role of a chemical potential shift, $\mu ^{\left(
1\right) }-\mu _{0}$, with respect to the non-interacting value $\mu _{0}.$
We use the notation $\mu ^{\left( 1\right) }$ with a superscript $\left(
1\right) $ to note that this is the chemical potential corresponding to the
self-energy $\Sigma ^{\left( 1\right) }.$ We will give additional supporting
evidence for this conjecture in section \ref{S_Exact} that discusses exact
results satisfied by our approach and in Appendix \ref{Annexe_regle_de_somme}%
. In actual calculations the chemical potential used in the Green function
occurring at this step of the calculation is determined from the condition $%
G_{11}^{\left( 1\right) }\left( 1,1^{+}\right) =\left\langle n_{\uparrow
}\left( 1\right) \right\rangle $ with 
\begin{equation}
G_{11}^{\left( 1\right) }\left( {\bf k},ik_{n}\right) =\frac{1}{%
ik_{n}-\varepsilon _{{\bf k}}+\mu ^{\left( 1\right) }-{\bf \Sigma }%
_{11}^{\left( 1\right) }\left( {\bf k},ik_{n}\right) }  \label{DefG(1)}
\end{equation}
${\bf \Sigma }_{11}^{\left( 1\right) }\left( {\bf k},ik_{n}\right) $ in the
combination $\mu ^{\left( 1\right) }-{\bf \Sigma }_{11}^{\left( 1\right)
}\left( {\bf k},ik_{n}\right) $ is a constant so that $\mu ^{\left( 1\right)
}-{\bf \Sigma }_{11}^{\left( 1\right) }\left( {\bf k},ik_{n}\right) $ equals
the non-interacting chemical potential $\mu _{0}$.

It is important to keep in mind the following. The approximate analytical
expression Eq.(\ref{Self_ordre_0_et_pot_chim}) for ${\bf \Sigma }^{\left(
1\right) }$ and the corresponding chemical potential $\mu ^{\left( 1\right)
} $ are useful for analytical arguments and numerical estimates\cite
{NoteAnneMarieU>0}. However, more fundamentally, the chemical potential is a
thermodynamic quantity that can be computed in a consistent way, along with
all other thermodynamic quantities. In diagrammatic methods, the
Luttinger-Ward functional provides a systematic method to obtain such
thermodynamically consistent results.\cite{LuttingerWard:1960} Since our
approach is non-perturbative, another approach must be used. At this point, $%
\mu ^{\left( 1\right) }$ is a first estimate for the true chemical
potential. A better estimate is $\mu ^{\left( 2\right) },$ corresponding to $%
{\bf \Sigma }^{\left( 2\right) }$, the next level of approximation for the
self-energy.

\subsection{ TPSC plus improved approximation for the self-energy}

\label{Sec_TPSC_self(2)}

Let us summarize what we have up to now. In this TPSC approach for the
attractive Hubbard model, there is only one particle-particle irreducible
vertex $U_{pp}$ Eq.(\ref{TPSC_vertex}). In addition, the two pair
susceptibilities that we need, namely 
\begin{eqnarray}
-\left. \frac{\delta G_{21}\left( 2,2;\left\{ {\bf \Theta }\right\} \right) 
}{\delta \theta ^{\ast }\left( 1,1\right) }\right| _{{\bf \Theta =0}}
&=&-\left\langle T_{\tau }\left[ \psi _{\downarrow }\left( 1\right) \psi
_{\uparrow }\left( 1\right) \psi _{\downarrow }^{\dagger }\left( 2\right)
\psi _{\uparrow }^{\dagger }\left( 2\right) \right] \right\rangle
\label{Suscep_espaceConf} \\
-\left. \frac{\delta G_{12}\left( 2,2;\left\{ {\bf \Theta }\right\} \right) 
}{\delta \theta \left( 1,1\right) }\right| _{{\bf \Theta =0}}
&=&-\left\langle T_{\tau }\left[ \psi _{\uparrow }^{\dagger }\left( 1\right)
\psi _{\downarrow }^{\dagger }\left( 1\right) \psi _{\uparrow }\left(
2\right) \psi _{\downarrow }\left( 2\right) \right] \right\rangle
\label{Suscep_espaceConf_b}
\end{eqnarray}
are quite easily related by the operation $2\leftrightarrow 1$, and
anticommutation $\psi _{\uparrow }\left( 2\right) \psi _{\downarrow }\left(
2\right) =-\psi _{\downarrow }\left( 2\right) \psi _{\uparrow }\left(
2\right) $. So we define 
\begin{equation}
\chi _{p}\left( 1,2\right) \equiv -\left. \frac{\delta G_{21}\left(
2,2;\left\{ {\bf \Theta }\right\} \right) }{\delta \theta ^{\ast }\left(
1,1\right) }\right| _{{\bf \Theta =0}}=-\left. \frac{\delta G_{12}\left(
1,1;\left\{ {\bf \Theta }\right\} \right) }{\delta \theta \left( 2,2\right) }%
\right| _{{\bf \Theta =0}}~.  \label{Suscep_chi_p}
\end{equation}
The equality of these two functions is a reflection of the fact that for
on-site pairing, the Pauli principle makes the triplet channel vanish.
Hence, substituting the particle-particle irreducible vertex Eq.(\ref
{TPSC_vertex}) in one of the Bethe-Salpeter equations Eq.(\ref{BetheSalTheta}%
) and expressing the result in terms of the ordinary Matsubara Green
functions 
\begin{eqnarray}
G_{\uparrow }^{\left( 1\right) }\left( 1,2\right) &=&G_{11}^{\left( 1\right)
}\left( 1,2\right) \\
G_{\downarrow }^{\left( 1\right) }\left( 1,2\right) &=&-G_{22}^{\left(
1\right) }\left( 2,1\right)
\end{eqnarray}
we find the pair susceptibility at the first level of approximation $\chi
_{p}^{\left( 1\right) }\left( 1,2\right) $, 
\begin{equation}
\chi _{p}^{\left( 1\right) }\left( 1,2\right) =G_{\uparrow }^{\left(
1\right) }\left( 1,2\right) G_{\downarrow }^{\left( 1\right) }\left(
1,2\right) -U_{pp}G_{\uparrow }^{\left( 1\right) }\left( 1,\overline{4}%
\right) \chi _{p}^{\left( 1\right) }\left( \overline{4},2\right)
G_{\downarrow }^{\left( 1\right) }\left( 1,\overline{4}\right) ~.
\label{BS1}
\end{equation}
In Fourier space and with the definitions, $q=\left( {\bf q},iq_{n}\right) ,$
$q_{n}=2n\pi T,$ $k=\left( {\bf k},ik_{n}\right) ,$ $k_{n}=\left(
2n+1\right) \pi T$ and 
\begin{equation}
\chi _{ir}^{\left( 1\right) }\left( q\right) =\frac{T}{N}\sum_{k}G_{%
\downarrow }^{\left( 1\right) }\left( k+q\right) G_{\uparrow }^{\left(
1\right) }\left( -k\right)  \label{Def_chi(1)}
\end{equation}
for the particle-particle susceptibility that is irreducible with respect to 
$U_{pp},$ the above equation will take the form 
\begin{equation}
\chi _{p}^{\left( 1\right) }\left( q\right) =\frac{\chi _{ir}^{\left(
1\right) }\left( q\right) }{1+U_{pp}\chi _{ir}^{\left( 1\right) }\left(
q\right) }  \label{Suscep_matriceT}
\end{equation}

To close the set of equations, we need to solve for the irreducible vertex 
\begin{equation}
U_{pp}=U\frac{\left\langle n_{\downarrow }\left( 1-n_{\uparrow }\right)
\right\rangle }{\left\langle n_{\downarrow }\right\rangle \left\langle
1-n_{\uparrow }\right\rangle }~.  \label{TPSC_Upp}
\end{equation}
That may be done by using either of the two exact limits $2\rightarrow 1^{+}$
and $2\rightarrow 1^{-}$ of Eqs.(\ref{Suscep_espaceConf})(\ref{Suscep_chi_p}%
) 
\begin{equation}
\chi _{p}^{\left( 1\right) }\left( 1,1^{+}\right) =\left\langle
n_{\downarrow }n_{\uparrow }\right\rangle  \label{Regle_de_Somme_nn}
\end{equation}
\begin{equation}
\chi _{p}^{\left( 1\right) }\left( 1,1^{-}\right) =1-n+\left\langle
n_{\downarrow }n_{\uparrow }\right\rangle  \label{Regle_de_Somme_1-n+nn}
\end{equation}
The left hand side of the latter two equations, that one could call
local-pair sum rules, can be transformed into the following two sum rules
when our approximation is used for the susceptibility 
\begin{equation}
\chi _{p}^{\left( 1\right) }\left( 1,1^{+}\right) =\left\langle
n_{\downarrow }n_{\uparrow }\right\rangle =\frac{T}{N}\sum_{q}\frac{\chi
_{ir}^{\left( 1\right) }\left( q\right) }{1+U_{pp}\chi _{ir}^{\left(
1\right) }\left( q\right) }e^{-iq_{n}0^{-}}  \label{TPSC_fermeture_1}
\end{equation}
\begin{equation}
\chi _{p}^{\left( 1\right) }\left( 1,1^{-}\right) =1-n+\left\langle
n_{\downarrow }n_{\uparrow }\right\rangle =\frac{T}{N}\sum_{q}\frac{\chi
_{ir}^{\left( 1\right) }\left( q\right) }{1+U_{pp}\chi _{ir}^{\left(
1\right) }\left( q\right) }e^{-iq_{n}0^{+}}  \label{TPSC_fermeture_2}
\end{equation}
We will discuss in section \ref{S_Exact} why, in our approach, both sum
rules are consistent and so give exactly the same result for $U_{pp}$. Monte
Carlo simulations confirm that the pair susceptibilities calculated with
Eqs.(\ref{TPSC_Upp}), (\ref{TPSC_fermeture_1}),(\ref{Def_chi(1)}) and (\ref
{DefG(1)}) are excellent approximations from weak to intermediate coupling.

As we saw above, at this stage the self-energy $\Sigma ^{\left( 1\right) }$
entering the calculation is a constant determined in such a way that we have
the proper filling. At high-frequency the exact limit of the self-energy is
the Hartree-Fock result and the corresponding irreducible vertex is the bare
interaction. In our section on exact results, we have found an expression
for the self-energy, Eqs.(\ref{Self_Exacte_HFexplicite11})(\ref
{Self_Exacte_HFexplicite22}), that is the sum of two terms, a first one
which is the high-frequency Hartree-Fock behavior and a second one that only
involves the low frequency behavior of Green functions, of irreducible
vertices and of susceptibilities. As we have seen above, our approximate
expressions for these quantities are consistent and are low-frequency
approximations. Hence, we substitute in Eqs.(\ref{Self_Exacte_HFexplicite11}%
)(\ref{Self_Exacte_HFexplicite22}) the result Eq.(\ref{BS1}) for the
susceptibility $-\left. \delta G_{21}^{\left( 1\right) }\left( 2,2;\left\{ 
{\bf \Theta }\right\} \right) /\delta \theta ^{\ast }\left( 1,1\right)
\right| _{{\bf \Theta =0}}=\chi _{p}^{\left( 1\right) }\left( 1,2\right) ,$
and the corresponding results Eq.(\ref{TPSC_vertex}) for the irreducible
vertex $\left. \delta \Sigma _{21}^{\left( 1\right) }\left( 1,2;\left\{ {\bf %
\Theta }\right\} \right) /\delta G_{21}^{\left( 1\right) }\left( 3,4;\left\{ 
{\bf \Theta }\right\} \right) \right| _{{\bf \Theta =0}}$ and Green function 
${\bf G}^{\left( 1\right) }.$ One is left with 
\begin{equation}
{\bf \Sigma }_{\uparrow }^{\left( 2\right) }\left( 1,2\right)
=UG_{\downarrow }^{\left( 1\right) }\left( 1,1^{+}\right) \delta \left(
1-2\right) -UU_{pp}G_{\downarrow }^{\left( 1\right) }\left( 2,1\right) \chi
_{p}^{\left( 1\right) }\left( 1,2\right)  \label{Sigma(1)_espace_reel}
\end{equation}
Appendix \ref{Annexe_SC-T_repulsive} discusses the relation to the
corresponding formula in the repulsive case. Note that all the terms on the
right-hand side of this equation, including the irreducible vertex $U_{pp}$,
are at the same level of approximation. In particular, the irreducible
vertex $U_{pp}$ that appears explicitly, is the functional derivative of the
field-dependent ${\bf \Sigma }^{\left( 1\right) }$ that enters $G^{\left(
1\right) }$ and $\chi _{p}^{\left( 1\right) }.$ This is crucial for the
quality of the approximation\cite{Vilk:1997,Moukouri:2000}. This type of
consistency is absent in many modern self-consistent treatments whose
self-energy contains renormalized Green's functions but with only bare
vertices.\cite{Bickers:1989} Going to Fourier space and using the full
expression for the susceptibility Eq.(\ref{Suscep_matriceT}), we have 
\begin{equation}
{\bf \Sigma }_{\uparrow }^{\left( 2\right) }\left( k\right) =Un_{\downarrow
}-U\frac{T}{N}\sum_{q}U_{pp}G_{\downarrow }^{\left( 1\right) }\left(
-k+q\right) \frac{\chi _{ir}^{\left( 1\right) }\left( q\right) }{%
1+U_{pp}\chi _{ir}^{\left( 1\right) }\left( q\right) }  \label{Self_(1)}
\end{equation}
Spin rotational invariance gives us the result for down spins. Note that one
of the vertices is bare while the other is dressed, contrary to the case
where there is a Migdal theorem.

The superscript $\left( 2\right) $ on the last expression for the
self-energy indicates that it is the next level of approximation. To improve
the susceptibility calculation we would need the irreducible vertices
corresponding to ${\bf \Sigma }_{12}^{\left( 2\right) }\left( 1,2;\left\{ 
{\bf \Theta }\right\} \right) $ which we do not have. Hence the calculation
stops at this level. Physically, the collective modes are less sensitive to
details of the quasiparticles, so they can be computed first with simple
Green functions. The self-energy on the other hand is sensitive to the
collective modes (they have zero-frequency Matsubara contributions contrary
to fermionic quantities) and hence we have to take these modes into account
when we want a better approximation for the self-energy.

\subsection{Internal accuracy check}

\label{Section_Consistency}

Either by returning to the derivation Eq.(\ref{Sig11etTr(SG)}) of ${\bf %
\Sigma }_{\uparrow }^{\left( 2\right) }\left( 1,2\right) $, which involves a
four point function Eq.(\ref{Suscep_espaceConf}) related to the pair
susceptibility Eqs.(\ref{BS1},\ref{Suscep_chi_p}) and double-occupancy Eq.(%
\ref{Regle_de_Somme_nn}), or by starting from the Fourier space expression
Eq.(\ref{Self_(1)}) and using the local pair sum rule Eq.(\ref
{TPSC_fermeture_1}) for $\chi _{p}^{\left( 1\right) }$ and the corresponding
sum rule for $\chi _{ir}^{\left( 1\right) },$ one finds that Eq.(\ref
{Sigma(1)_espace_reel}) satisfies 
\begin{equation}
\Sigma _{\uparrow }^{\left( 2\right) }\left( 1,\overline{2}\right)
G_{\uparrow }^{\left( 1\right) }\left( \overline{2},1^{+}\right) =\frac{T}{N}%
\sum_{k}\Sigma _{\uparrow }^{\left( 2\right) }\left( k\right) G_{\uparrow
}^{\left( 1\right) }\left( k\right) e^{-ik_{n}0^{-}}=U\left\langle
n_{\downarrow }n_{\uparrow }\right\rangle ~.  \label{Coherence_occ_doub}
\end{equation}
$U\left\langle n_{\downarrow }n_{\uparrow }\right\rangle $ entering this
equation is exactly the same as that computed from the local pair sum rule
Eq.(\ref{TPSC_fermeture_1}). This result is analogous to that found in the
repulsive model. Note that $G_{\uparrow }^{\left( 1\right) }$ enters the
above equation. An indication of the accuracy of our approximations may be
obtained by checking by how much $\Sigma _{\uparrow }^{\left( 2\right)
}\left( 1,\overline{2}\right) G_{\uparrow }^{\left( 2\right) }\left( 
\overline{2},1^{+}\right) $ differs from $U\left\langle n_{\downarrow
}n_{\uparrow }\right\rangle $ obtained from the local pair sum rule. We have
checked that there is at most a few percent discrepancy between both
calculations, except in the pseudogap regime. Note that the chemical
potential $\mu ^{\left( 2\right) }$ entering $G_{\uparrow }^{\left( 2\right)
}$ must be obtained from the number conservation equation. We refer to Ref. 
\cite{vilk1_322} for a discussion of Luttinger's theorem in this context.

In any approximation for the many-body problem, full consistency for all
correlation functions is unachievable. For example, in conserving
approximations\cite{Baym:1962,Bickers:1989} one starts from diagrams for the
Luttinger-Ward functional and for the corresponding free energy. Then a
self-energy and irreducible vertices are obtained from functional
differentiation. These quantities may then be used in the Bethe-Salpeter
equation to obtain the pair susceptibility. From that pair susceptibility,
one can compute double occupancy through the exact result Eq.(\ref
{Regle_de_Somme_nn}). The latter double occupancy is in general different
from the one obtained from $\Sigma \left( 1,\overline{2}\right) G\left( 
\overline{2},1^{+}\right) $ since it does not contain the same set of
diagrams. So if the double occupancy obtained from the susceptibility Eq.(%
\ref{Regle_de_Somme_nn}) is integrated over coupling constant to obtain the
free energy, the result will in general be different from the original free
energy. Conserving approximations are not self-consistent at the
two-particle level. Other criticisms of these approaches appear in Refs.\cite
{Vilk:1997} and \cite{Moukouri:2000}.

\section{Exact results satisfied by our non-perturbative approach}

\label{S_Exact}

We briefly discuss exact relations and consistency requirements that are
satisfied by our approach. Details of some of the proofs may be found in
Appendix \ref{Annexe_regle_de_somme}.

\subsection{Sum rules on single-particle spectral weight}

Consider first the single-particle properties. These should be calculated
with ${\bf G}^{\left( 2\right) }$ that contains the self-energy Eq.(\ref
{Self_(1)}) and the corresponding chemical potential. One can extract the
moments of the corresponding spectral weight ${\bf A}^{\left( 2\right)
}\left( {\bf k,}\omega \right) $ from the high-frequency expansion of ${\bf G%
}^{\left( 2\right) }\left( {\bf k},ik_{n}\right) $ in analogy with Eq.(\ref
{G(k,w)_haute_frequence}). In the self-energy $\Sigma ^{\left( 2\right)
}\left( {\bf k},ik_{n}\right) $, Eq.(\ref{Self_(1)}), the Hartree-Fock
contribution appears explicitly so that the exact high-frequency limit Eq.(%
\ref{SigmaHF}) is satisfied. This means that the normalization and first
moment of ${\bf A}^{\left( 2\right) }\left( {\bf k,}\omega \right) $ satisfy
the exact results Eq.(\ref{A(k,w)_normalisation}) and Eq.(\ref
{A(k,w)_premier_moment}).

\subsection{Sum rules on pair spectral weight}

Concerning two-particle properties, more specifically the pairing
susceptibility, there are a number of exact results that our approach
satisfies. First there are the two local-pair sum rules Eqs.(\ref
{TPSC_fermeture_1})(\ref{TPSC_fermeture_2}) that are a consequence of the
fluctuation-dissipation theorem. The value of $U_{pp}$ obtained from either
one of them is the same. This is demonstrated in Appendix \ref
{Annexe_regle_de_somme}.

The other exact properties we shall consider concern the pair spectral
weight. Using the definition of the pair field $\Delta $ Eq.(\ref{Def_Delta}%
) and of the pair susceptibility Eqs.(\ref{Suscep_chi_p})(\ref
{Suscep_espaceConf})(\ref{Suscep_espaceConf_b}) the latter may be written as 
\begin{equation}
\chi _{ex}\left( 1,2\right) =\left\langle T_{\tau }\Delta \left( 1\right)
\Delta ^{\dagger }\left( 2\right) \right\rangle .
\end{equation}
The subscript $ex$ stresses that we are, for now, considering properties of
the exact pair susceptibility. The Lehmann representation, and the
periodicity in imaginary time, allow one to show that 
\begin{equation}
\chi _{ex}\left( {\bf q},iq_{n}\right) =\int \frac{d\omega }{\pi }\frac{\chi
_{ex}^{\prime \prime }\left( {\bf q},\omega \right) }{\omega -iq_{n}}
\label{Repr_spectrale}
\end{equation}
where the time Fourier transform of the pair spectral weight $\chi
_{ex}^{\prime \prime }\left( {\bf q},\omega \right) $ is defined by 
\begin{equation}
\chi _{ex}^{\prime \prime }\left( {\bf q},t\right) =\frac{1}{2}\left\langle %
\left[ \Delta _{{\bf q}}\left( t\right) ,\Delta _{{\bf q}}^{\dagger }\left(
0\right) \right] \right\rangle ~.
\end{equation}
From the latter definition and the equations of motion Eqs.(\ref{EqMvtDelta}%
)(\ref{EqMvtDelta+}) one can show that the quantity $\chi _{ex}^{\prime
\prime }\left( {\bf q},\omega \right) $ obeys the following sum-rule 
\begin{equation}
\int \frac{d\omega }{\pi }\chi _{ex}^{\prime \prime }\left( {\bf q},\omega
\right) =\left\langle \left[ \Delta _{{\bf q}}\left( 0\right) ,\Delta _{{\bf %
q}}^{\dagger }\left( 0\right) \right] \right\rangle =1-n  \label{Moment_0}
\end{equation}
where $n$ is the filling that is obtained from the single-particle Green
functions entering the calculation of $\chi _{ex}^{\prime \prime }\left( 
{\bf q},\omega \right) $. We show in Appendix \ref{Annexe_regle_de_somme}
that our approximate expression for the susceptibility Eq.(\ref
{Suscep_matriceT}) satisfies this manifestation of the Pauli principle
exactly, for all wave vectors ${\bf q}$. That is why we can use either of
the two local pair sum-rules Eqs.(\ref{TPSC_fermeture_1})(\ref
{TPSC_fermeture_2}) to find self-consistently the value of $\left\langle
n_{\uparrow }n_{\downarrow }\right\rangle .$

Proceeding to the first moment of the pair spectral weight, it is shown in
Appendix \ref{Annexe_regle_de_somme}, that 
\begin{eqnarray}
\int \frac{d\omega }{\pi }\omega \chi _{ex}^{\prime \prime }\left( {\bf q}%
,\omega \right) &=&\frac{1}{N}\sum_{{\bf k}}\left( \varepsilon _{{\bf k}%
}+\varepsilon _{-{\bf k+q}}-2\left( \mu -\frac{U}{2}\right) \right) \left(
1-\left\langle n_{{\bf k\uparrow }}\right\rangle -\left\langle n_{-{\bf %
k+q\downarrow }}\right\rangle \right) \\
&=&\left[ \frac{1}{N}\sum_{{\bf k}}\left( \varepsilon _{{\bf k}}+\varepsilon
_{-{\bf k+q}}\right) \left( 1-2\left\langle n_{{\bf k\uparrow }%
}\right\rangle \right) \right] -2\left( \mu -\frac{U}{2}\right) \left(
1-n\right) ~.  \label{Chi_regle_de_somme_f}
\end{eqnarray}
Like the previous sum rule, this result is valid for {\it all }wave vectors $%
{\bf q}.$ It is a generalization of the $f-$sum rule to the case of the
attractive Hubbard model, a sort of off-diagonal $f-$sum rule. It is
generalized Ward identity that relates two-particle quantities on the
left-hand side with quantities obtained from the one-particle Green function
on the right-hand side. At half-filling, $\mu -\frac{U}{2}=0$, where there
is an exact canonical transformation from the attractive to the repulsive
Hubbard model, the above result reduces precisely to the $f-$sum-rule for
the repulsive case. Again, the above expression Eq.(\ref
{Chi_regle_de_somme_f}) relates a two-particle quantity, on the left-hand
side, to a single particle property, on the right-hand side. Neither the
pair susceptibility nor the single-particle Green function are known exactly
in our approach. Nevertheless, as shown in Appendix \ref
{Annexe_regle_de_somme}, when our approximation $\chi _{p}^{\left( 1\right)
} $ for the pair susceptibility is substituted on the left-hand side of Eq.(%
\ref{Chi_regle_de_somme_f}), and our expression for the corresponding
single-particle Green function ${\bf G}^{\left( 1\right) }$ is substituted
on the right-hand side, the equation is satisfied exactly as long as one
uses 
\begin{equation}
\mu ^{\left( 1\right) }=\mu _{0}+\frac{U}{2}-\frac{U_{pp}}{2}\left(
1-n\right)  \label{Chi_somme_f_et_mu}
\end{equation}
for the chemical potential appearing on the right-hand side. This is
consistent with the fact that the quantities entering the right-hand side of
this equation must pertain to the single-particle Green functions used in
the calculation of $\chi _{p}^{\prime \prime }\left( {\bf q},\omega \right) $
on the left-hand side. Since the chemical potential entering $G^{\left(
1\right) }$ should be $\mu ^{\left( 1\right) }=\mu _{0}+$ $\Sigma ^{\left(
1\right) }$ to obtain the correct filling, use of the approximation Eq.(\ref
{Self_ordre_0_et_pot_chim}) for $\Sigma ^{\left( 1\right) }$ leads to the
above result Eq.(\ref{Chi_somme_f_et_mu}). Recall however that Eq.(\ref
{Self_ordre_0_et_pot_chim}) for $\Sigma ^{\left( 1\right) }$ is not
rigorous. Nevertheless, away from the renormalized classical regime, where
the self-energy is weakly frequency dependent, the chemical potential
obtained from this formula differs little from the one obtained at the
second level of approximation $\Sigma ^{\left( 2\right) }$ or from Monte
Carlo simulations in the weak to intermediate coupling regime\cite
{KyungAllen:2000}. The result Eq.(\ref{Chi_regle_de_somme_f}) can also be
considered as a type of consistency condition between one and two-particle
quantities analogous to the relation between ${\rm Tr}\left( \Sigma G\right) 
$ and $U\left\langle n_{\downarrow }n_{\uparrow }\right\rangle $ discussed
in section \ref{Section_Consistency}.

\subsection{Miscellaneous}

The Mermin-Wagner theorem is satisfied by our approach. That theorem states
that classical fluctuation effects prohibit continuous symmetry breaking in
two dimensions. The proof follows the steps of Appendix A.3 Ref.\cite
{Vilk:1997}. The Kosterlitz-Thouless-Berezinskii\cite{KTB} transition on the
other hand involves algebraic order and large-scale vortex structures that
are absent from the present approach. This transition is thus inaccessible
to us. Far from the critical point on the other hand, one can show,
following the steps Ref.\cite{vilk1e} in analogy with the repulsive case,
that there is Kanamori-Brueckner (quantum-fluctuation) type screening of $%
U_{pp}$ which is given by the approximate formula 
\begin{equation}
U_{pp}\simeq \frac{U}{1-\Lambda U}\quad ;\quad \Lambda =\frac{T}{N}\frac{1}{%
\left\langle n_{{\bf \uparrow }}\right\rangle \left\langle 1-n_{{\bf %
\downarrow }}\right\rangle }\sum_{q}\left[ \chi _{ir}^{\left( 1\right)
}\left( q\right) \right] ^{2}~.
\end{equation}

Finally, it is important to notice that all calculations are done at
constant density. In particular, the Green function $G^{\left( 1\right) }$
entering the calculation of $\Sigma ^{\left( 2\right) }$ in Eq.(\ref
{Self_(1)}) is evaluated at the same density as the final result $G^{\left(
2\right) }$. This is motivated by the existence of Luttinger's theorem which
states that the volume enclosed by the Fermi surface at $T=0$ depends only
on density, not on interaction. It would be unphysical to iterate from $%
\Sigma ^{\left( 1\right) }$ to $\Sigma ^{\left( 2\right) }$ starting from a $%
G^{\left( 1\right) }$ whose Fermi-surface associated singularities are at
locations in the Brillouin zone that never intersect those of $G^{\left(
2\right) }$. The constant-density constraint ensures maximum overlap. This
point of view is motivated by Luttinger's approach. It is discussed further
in Ref.\cite{VilkEq47:1997}. Luttinger's theorem should be satisfied to a
very good approximation in our approach, as for the repulsive model\cite
{Vilk:1997}.

\section{Conclusion}

\label{Section_conclusion}

In this paper we have presented a generalization of the approach developed
in Ref.\cite{Vilk:1997} to the attractive Hubbard model. We first
established a number of exact results that form the basis of the
approximation method that we introduced in section \ref{Section_TPSC}. The
first level of approximation (Two-Particle Self-Consistent) is based on a
Hartree-Fock like factorization {\it ansatz} for the self-energy in the
presence of an external off-diagonal field, Eq.(\ref{TPSC_Factorise}). That 
{\it ansatz} differs from the standard Hartree-Fock factorization Eq.(\ref
{SigmaG_factoriseHF}) by the presence of a constant matrix ${\bf A}\left(
\Theta \right) $ that forces the factorization {\it ansatz} to reduce to the
unfactorized four-point function in the special case where the later
involves only density-density correlations (double occupancy). That {\it %
ansatz} leads to the irreducible vertex for pair fluctuations simply through
functional differentiation. The resulting irreducible vertex, given by Eq.(%
\ref{TPSC_Upp}), depends on double-occupancy, a quantity that may then be
determined self-consistently using fluctuation-dissipation theorem derived
sum rules Eqs.(\ref{TPSC_fermeture_1}) and (\ref{TPSC_fermeture_2}). Either
one of these local-pair sum rules suffices to close the system of equations
since they are equivalent. That exact equivalence is satisfied in our
approach because the normalization sum-rule for the pair spectral weight,
Eq.(\ref{Moment_0}), is obeyed. This sum rule is a manifestation of the
Pauli principle.

The self-energy ${\bf \Sigma }^{\left( 1\right) }$ entering the
single-particle Green function at that first level of approximation is
constant. The value of this constant is irrelevant for the calculation of
the pair susceptibility since we work at constant filling, which means ${\bf %
\Sigma }^{\left( 1\right) }$ can be absorbed in the chemical potential.
Nevertheless, we have argued that Eq.(\ref{Self_ordre_0_et_pot_chim}) should
be a good approximation for the value of the constant self-energy at that
first level of approximation since it follows from a consistency requirement
between the {\it ansatz} Eq.(\ref{TPSC_Factorise}) and the two possible
values of ${\rm Tr}\left( \Sigma G\right) .$ In addition, Eq.(\ref
{Self_ordre_0_et_pot_chim}) for ${\bf \Sigma }^{\left( 1\right) }$ formally
allows the first-moment $\left( \int \frac{d\omega }{\pi }\omega \chi
_{p}^{\prime \prime }\left( {\bf q},\omega \right) \right) $ sum rule on the
pair spectral weight to be satisfied (Appendix \ref{Annexe_regle_de_somme}).
That sum rule is the off-diagonal generalization of the $f-$ sum rule
familiar from the particle-hole channel.

The rough approximation that the self-energy is a constant suffices to
obtain a good approximation for the low-frequency pair susceptibility since
collective modes do not generally depend strongly on details of the damping
of the underlying fermions. Details of single-particle damping do however
depend strongly on the collective modes. It is possible then to improve our
approximation for the self-energy by using our results for the collective
modes in an exact formula Eq.(\ref{Self_Exacte_HFexplicite11}) for $\Sigma $
where the high-frequency limit appears explicitely. That gives us a better
(cooperon-type) approximation for the self-energy Eq.(\ref{Self_(1)}).
Clearly, that approximation does not assume that Migdal's theorem is
satisfied since one of the vertices is dressed. When Migdal's theorem
applies, both vertices are bare. As in previous work, \cite{Vilk:1997}, one
can use the difference between $\Sigma _{\uparrow }^{\left( 2\right) }\left(
1,\overline{2}\right) G_{\uparrow }^{\left( 2\right) }\left( \overline{2}%
,1^{+}\right) $ and $U\left\langle n_{\downarrow }n_{\uparrow }\right\rangle 
$ as a check on the level of accuracy of approach, as discussed in Sec.\ref
{Section_Consistency}. Formal comparisons with other approaches are
presented in Appendix \ref{Annexe_comparaison}. Note that our approach is in
the $SO\left( N\rightarrow \infty \right) $ universality class, by arguments
similar to those that apply to the repulsive model.\cite{Dare} Hence, the
Mermin-Wagner theorem is satisfied as it should, but algebraic long-range
order of the Kosterlitz-Thouless type is beyond the accuracy of any
microscopic theory that does not use renormalization group arguments to
reach the long wavelength limit and treat the $SO\left( 2\right) $ symmetry
exactly.

In summary then, in its simplest form, our generalization of Ref.\cite
{Vilk:1997} to the attractive Hubbard model is expressed by the three simple
equations, Eqs.(\ref{TPSC_Upp}),(\ref{TPSC_fermeture_1}) and (\ref{Self_(1)}%
) plus the constant density contraint that determines the chemical potential
appropriate to the level of approximation. Extensions of this approach have
also been proposed.\cite{Allen:2000b} Questions related to thermodynamic
consistency and calculation of other response functions will be presented in
a later publication. In the accompanying paper, our approach is compared in
detail with results of Quantum Monte Carlo simulations. In addition to
achieving quantitative agreement with simulations, this approach predicts
the appearance of a pseudogap in the single-particle spectral weight when
the pair fluctuations enter the renormalized classical regime. This is
qualitatively different from the results obtained from self-consistent $%
T-matrix$ approximation, or FLEX-type approaches. The role of the low space
dimension, the pair correlation length and single-particle thermal de
Broglie wave length, and more generally the mechanism for the opening of
this pseudogap, have been discussed in detail in Ref.\cite{Vilk:1997}.

\acknowledgements

A.-M.S.T is indebted to Y.M. Vilk for numerous invaluable discussions. We
are also indebted to Bumsoo Kyung for calculations that allowed the results
of this work to be compared with Monte Carlo calculations, as well as for
corrections and critical reading of the manuscript. This work was partially
supported by the Natural Sciences and Engineering Research Council of Canada
(NSERC), by the Fonds pour la Formation de Chercheurs et l'Aide \`{a} la
Recherche (FCAR) from the Qu\'{e}bec government and by the Canadian
Institute for Advanced Research.

%TCIMACRO{
%\TeXButton{Appendix}{\appendix%
%}}%
%BeginExpansion
\appendix%
%
%EndExpansion

\section{Sum rules and high-frequency expansion for the pair susceptibility}

\label{Annexe_regle_de_somme}

In this appendix we give more details on the derivation of the results
presented in section \ref{S_Exact}. Let us begin with the high-frequency
expansion of the pair susceptibility $\chi _{p}\left( {\bf q},iq_{n}\right) $
for any approximation that has a spectral representation such as Eq.(\ref
{Repr_spectrale})$.$ From that spectral representation, one obtains, 
\begin{equation}
\chi _{p}\left( {\bf q},iq_{n}\right) \approx -\int \frac{d\omega }{\pi }%
\chi _{p}^{\prime \prime }\left( {\bf q},\omega \right) \left( \frac{1}{%
iq_{n}}\right) -\int \frac{d\omega }{\pi }\omega \chi _{p}^{\prime \prime
}\left( {\bf q},\omega \right) \left( \frac{1}{iq_{n}}\right) ^{2}+\ldots
\label{Dev_haute_frequ}
\end{equation}
Let us now consider the exact susceptibility $\chi _{ex}\left( {\bf q}%
,iq_{n}\right) $. The moments of $\chi _{ex}^{\prime \prime }\left( {\bf q}%
,\omega \right) $ that appear as coefficients of the expansion in powers of $%
\left( iq_{n}\right) ^{-1}$ may be obtained as follows. The first one
follows for all values of ${\bf q}$ from the equal-time commutator in Eq.(%
\ref{Moment_0}). The next coefficient, $\int \frac{d\omega }{\pi }\omega
\chi _{ex}^{\prime \prime }\left( {\bf q},\omega \right) $, follows from 
\begin{eqnarray}
\int \frac{d\omega }{\pi }\omega \chi _{ex}^{\prime \prime }\left( {\bf q}%
,\omega \right) &=&i\left[ \frac{\partial }{\partial t}\int \frac{d\omega }{%
\pi }e^{-i\omega t}\chi _{ex}^{\prime \prime }\left( {\bf q},\omega \right) %
\right] _{t=0} \\
&=&i\left[ \left\langle \frac{\partial }{\partial t}\left[ \Delta _{{\bf q}%
}\left( t\right) ,\Delta _{{\bf q}}^{\dagger }\left( 0\right) \right]
\right\rangle \right] _{t=0} \\
&=&-\left[ \left\langle \left[ \frac{\partial }{\partial \tau }\Delta _{{\bf %
q}}\left( \tau \right) ,\Delta _{{\bf q}}^{\dagger }\left( 0\right) \right]
\right\rangle \right] _{\tau =0}~.
\end{eqnarray}
The latter equal-time commutator may be computed after $\frac{\partial }{%
\partial \tau }\Delta _{{\bf q}}\left( \tau \right) $ is rewritten with the
help of the equation of motion, Eq.(\ref{EqMvtDelta}), leading to the result
Eq.(\ref{Chi_regle_de_somme_f}).

Since our approximate expression Eq.(\ref{Suscep_matriceT}) for the pair
susceptibility admits a spectral representation, its moments may be obtained
from the high-frequency expansion in powers of $\left( iq_{n}\right) ^{-1}$,
by analogy with the method in Ref.\cite{Vilk:1997}. From our approximate
formula Eq.(\ref{Suscep_matriceT}) $\chi _{p}^{\left( 1\right) }\left(
q\right) ^{-1}=\chi _{ir}^{\left( 1\right) }\left( q\right) ^{-1}+U_{pp}$
and the large $iq_{n}$ expansion of $\chi _{ir}^{\left( 1\right) }\left(
q\right) $ one finds that 
\begin{equation}
\lim_{iq_{n}\rightarrow \infty }iq_{n}\chi _{p}^{\left( 1\right) }\left( 
{\bf q},iq_{n}\right) =\lim_{iq_{n}\rightarrow \infty }iq_{n}\chi
_{ir}^{\left( 1\right) }\left( {\bf q},iq_{n}\right) =-\left( 1-n\right)
\label{Coefficient_iqn-1}
\end{equation}
in agreement with the exact result Eq.(\ref{Moment_0}). Note that the large $%
iq_{n}$ limit of $\chi _{ir}^{\left( 1\right) }\left( {\bf q},iq_{n}\right) $
must be taken after the sum over fermionic frequencies in Eq.(\ref
{Def_chi(1)}).

The first moment (off-diagonal $f$ sum-rule) is given by 
\begin{eqnarray}
-\int \frac{d\omega }{\pi }\omega \chi _{p}^{\prime \prime \left( 1\right)
}\left( {\bf q},\omega \right) &=&\lim_{iq_{n}\rightarrow \infty }\left[
\left( iq_{n}\right) ^{2}\chi _{p}^{\left( 1\right) }\left( {\bf q}%
,iq_{n}\right) +iq_{n}\left( 1-n\right) \right]  \nonumber \\
&=&\lim_{iq_{n}\rightarrow \infty }\left[ \left( iq_{n}\right) ^{2}\chi
_{ir}^{\left( 1\right) }\left( {\bf q},iq_{n}\right) \left( 1-U_{pp}\chi
_{ir}^{\left( 1\right) }\left( {\bf q},iq_{n}\right) \right) +iq_{n}\left(
1-n\right) \right]
\end{eqnarray}
Substituting the exact results for the high-frequency expansion of $\chi
_{ir}^{\left( 1\right) }\left( {\bf q},iq_{n}\right) ,$ one obtains, 
\begin{equation}
\int \frac{d\omega }{\pi }\omega \chi _{p}^{\prime \prime \left( 1\right)
}\left( {\bf q},\omega \right) =\frac{1}{N}\sum_{{\bf k}}\left( \varepsilon
_{{\bf k}}+\varepsilon _{-{\bf k+q}}\right) \left( 1-2\left\langle n_{{\bf %
k\uparrow }}\right\rangle \right) -2\mu _{0}\left( 1-n\right) +U_{pp}\left(
1-n\right) ^{2}  \label{Regle_de_somme_f_notre_approx}
\end{equation}
In this expression, $\left\langle n_{{\bf k\uparrow }}\right\rangle $ is
computed from $G^{\left( 1\right) },$ hence is a Fermi function, and $\mu
_{0}=\mu ^{\left( 1\right) }+\Sigma ^{\left( 1\right) }$ enters the Green
function $G^{\left( 1\right) }$ from which $\chi _{ir}^{\left( 1\right) }$
is computed$.$ Since the self-energy $\Sigma ^{\left( 1\right) }$ entering $%
G^{\left( 1\right) }$ is a constant, $\mu _{0}$ coincides with the
non-interacting chemical potential appropriate for the filling we are
considering. Comparing with the exact result Eq.(\ref{Chi_regle_de_somme_f})
we see that the chemical potential at that level of approximation should be
given by $\mu ^{\left( 1\right) }=\mu _{0}+\left( U-U_{pp}\left( 1-n\right)
\right) /2$ which coincides with our proposed approximation Eq.(\ref
{Self_ordre_0_et_pot_chim}). Numerically, we have checked that this
approximate chemical potential is quite close to the chemical potential $\mu
^{\left( 2\right) }$ obtained with $\Sigma ^{\left( 2\right) }$ and that the
latter in turn is close to those obtained from Monte Carlo simulations\cite
{KyungAllen:2000}, as long as we are far from the renormalized classical
regime where ${\bf \Sigma }^{\left( 2\right) }$ acquires a strong frequency
dependence.

\section{Comparison with other approaches}

\label{Annexe_comparaison}

We first comment on the connection to the formalism for the repulsive model.
This leads us then to a discussion of the widely used Self-Consistent $%
T-matrix$ approximation and of another ``mixed'' approach that has been
extensively applied recently.\cite{LevinTous}

\subsection{Mapping to the repulsive model and comparisons with
Self-Consistent $T-$matrix approximation}

\label{Annexe_SC-T_repulsive}

At half-filling, the Lieb-Mattis canonical transformation 
\begin{eqnarray}
c_{i,\uparrow } &\rightarrow &c_{i,\uparrow } \\
c_{i,\downarrow } &\rightarrow &e^{i{\bf Q\cdot r}_{i}}c_{i,\downarrow
}^{\dagger }
\end{eqnarray}
with ${\bf Q=}\left( \pi ,\pi \right) $ maps the repulsive into the
attractive Hubbard model. The same canonical transformation maps spins $S,$
density $\rho $ and pairing operators into each other as follows 
\begin{eqnarray}
S^{z}\left( {\bf Q+q}\right) &\rightarrow &\rho \left( {\bf Q+q}\right) 
\nonumber \\
S^{+}\left( {\bf Q+q}\right) &\rightarrow &-\Delta ^{\dagger }\left( {\bf q}%
\right)  \nonumber \\
S^{-}\left( {\bf Q+q}\right) &\rightarrow &-\Delta \left( {\bf q}\right)
\end{eqnarray}
For a chemical potential different from $U/2$ (half-filling), the repulsive
model maps into the attractive model at half-filling but in a finite
Zeeman-coupled magnetic field. The approach presented here for the
attractive model would, at half-filling, translate into the
transverse-channel calculation for the repulsive model\cite{Moukouri:2000}.
In that $U>0$ case, Ref.\cite{Vilk:1997_App_C} presents only the
longitudinal channel calculation. The analog calculation for the attractive
Hubbard model would have lead us to two irreducible vertices. One vertex
would have been for the charge fluctuations. These are related to pair
fluctuations by the $SO\left( 3\right) $ symmetry of the model a $n=1,$
which implies that the corresponding irreducible vertex there is also $%
U_{pp}.$ The other vertex would have been for the non-singular
spin-fluctuation channel. The two vertices would have appeared in a
self-energy formula that would replace Eq.(\ref{Sigma(1)_espace_reel}).
However, at $n=1$, the best self-energy formula for the attractive Hubbard
model would be obtained from the canonical transformation of that presented
in Ref.\cite{Moukouri:2000} which preserves crossing symmetry. For problems
sufficiently far away from half-filling $\left( T_{X}\ll \mu \right) $
however, the pair fluctuations suffice.\cite{NoteTx}

In the repulsive model case, we have presented general analytical arguments 
\cite{Vilk:1997} as well as detailed comparisons between: Monte Carlo
simulations \cite{Vilk:1997,Moukouri:2000}, our approach, and
self-consistent Eliashberg type approaches (such as the Fluctuation-Exchange
approximation). Most of our general criticism concerning self-consistent
approaches in the repulsive case apply to self-consistent $T-matrix$ plus
fluctuation exchange approaches in the attractive case.

More specifically, one of the key qualitative differences between our
approach and self-consistent approaches is that the latter do not predict
the existence of a fluctuation-induced pseudogap in the single-particle
spectral weight in two dimensions. We have demonstrated at length, through
comparisons with Monte Carlo simulations\cite{Moukouri:2000,Allen:1999} and
with physical arguments\cite{Vilk:1997}, that this is incorrect in both the
repulsive and the attractive cases.

\subsection{The $GG^{\left( 0\right) }$ approach}

An alternate approach based on computing the irreducible susceptibility with
one bare and one dressed Green function has been extensively used lately\cite
{LevinTous}. More specifically, in this approach 
\begin{equation}
\chi _{p}=\frac{\widetilde{\chi }_{p}}{1+U\widetilde{\chi }_{p}}
\label{Levin_suscep_tot}
\end{equation}
where 
\begin{equation}
\widetilde{\chi }_{p}=\frac{T}{N}\sum_{{\bf k}}G_{\downarrow }\left( {\bf k}%
\right) G_{\uparrow }^{0}\left( -{\bf k+q}\right)  \label{Levin_suscep}
\end{equation}
with the self-energy entering $G$ given by 
\begin{eqnarray}
\Sigma _{\downarrow }\left( k\right) &=&\frac{T}{N}\sum_{{\bf q}}\frac{U}{1+U%
\widetilde{\chi }_{p}\left( q\right) }G_{\uparrow }^{0}\left( q-k\right)
e^{-i\left( q_{n}-k_{n}\right) 0^{-}}  \label{Levin_self_T} \\
&=&Un_{\uparrow }^{0}-U^{2}\frac{T}{N}\sum_{{\bf q}}\frac{\widetilde{\chi }%
_{p}\left( q\right) }{1+U\widetilde{\chi }_{p}\left( q\right) }G_{\uparrow
}^{0}\left( q-k\right) .  \label{Levin_self_T_HF}
\end{eqnarray}
Since there is a single chemical potential for the two Green functions $%
G_{\downarrow }$ and $G_{\uparrow }^{0},$ there are two different
expressions for the occupation numbers operators ($n_{{\bf k}}$ and $n_{{\bf %
k}}^{0})$ and for the corresponding fillings, ($n$ and $n^{0}$)$.$

A positive aspect of this approach is that it exhibits consistency between
one and two particle properties in the sense that the exact result 
\begin{equation}
\frac{T}{N}\sum_{k}\Sigma _{\downarrow }\left( k\right) G_{\downarrow
}\left( k\right) e^{-ik_{n}0^{-}}=U\left\langle n_{\downarrow }n_{\uparrow
}\right\rangle
\end{equation}
that follows from the equation of motion, is satisfied exactly by the above
approximate scheme. Indeed, starting from the approximate formula Eq.(\ref
{Levin_self_T}) for the self-energy, and using Eq.(\ref{Levin_suscep}) for
the susceptibility, one finds that 
\begin{eqnarray}
\frac{T}{N}\sum_{k}\Sigma _{\downarrow }\left( k\right) G_{\downarrow
}\left( k\right) e^{-ik_{n}0^{-}} &=&U\frac{T}{N}\sum_{{\bf q}}\frac{%
\widetilde{\chi }_{p}\left( q\right) }{1+U\widetilde{\chi }_{p}\left(
q\right) }e^{-iq_{n}0^{-}} \\
&=&U\left\langle \Delta ^{\dagger }\Delta \right\rangle =U\left\langle
n_{\uparrow }n_{\downarrow }\right\rangle
\end{eqnarray}
One goes from the second to the last line using the fluctuation
dissipation-theorem. Hence, in this approach, the double occupancy obtained
from single particle quantities (namely $\Sigma _{\downarrow }G_{\downarrow
} $) is exactly the same as that found from the pair susceptibility, a
two-particle quantity.

On the negative side, the spectral weight corresponding to the
susceptibility Eq.(\ref{Levin_suscep_tot}) does not satisfy the sum rules on
the first two moments discussed in Eqs.(\ref{Chi_regle_de_somme_f}) and (\ref
{Moment_0}). To show this, we first need a few sum rules on the
single-particle spectral weight. Following the steps in Appendix A of Ref. 
\cite{Vilk:1997}, the high-frequency expansion of $G$, with $\Sigma $ given
by Eq.(\ref{Levin_self_T_HF}), gives the following sum rules for the
corresponding spectral weight $A\left( {\bf k,}\omega \right) $ 
\begin{equation}
\int \frac{d\omega }{2\pi }A\left( {\bf k},\omega \right) =1
\label{Normalisation}
\end{equation}
\begin{equation}
\int \frac{d\omega }{2\pi }\omega A\left( {\bf k},\omega \right)
=\varepsilon _{{\bf k}}-\mu +Un_{-\sigma }^{0}~.  \label{Premier_moment}
\end{equation}
Also, the equation of motion for $G$ gives 
\begin{equation}
\int \frac{d\omega }{2\pi }\omega f\left( \omega \right) A\left( {\bf k}%
,\omega \right) =\frac{1}{N}\sum_{{\bf k}}\left( \varepsilon _{{\bf k}}-\mu
\right) n_{{\bf k,\sigma }}+U\left\langle n_{\uparrow }n_{\downarrow
}\right\rangle ~.  \label{Omega_f_A}
\end{equation}
The above sum rules are valid for both $G$ and $G^{\left( 0\right) }.$ In
the latter case however, we take the Fermi function for the occupation
number and $U=0$ on the right-hand side of the above equations.

We are now ready to check the sum rules for $\widetilde{\chi }_{p}.$ Using
the spectral representation for $G$ in the expression for the susceptibility
Eq.(\ref{Levin_suscep}), 
\begin{equation}
\widetilde{\chi }_{p}\left( {\bf q},iq_{n}\right) =\frac{T}{N}\sum_{{\bf k}%
}\sum_{ik_{n}}\int \frac{d\omega }{2\pi }\frac{A\left( {\bf k,}\omega
\right) }{ik_{n}-\omega }\frac{1}{-ik_{n}+iq_{n}-\left( \varepsilon _{-{\bf %
k+q}}-\mu \right) }
\end{equation}
and performing the high-frequency expansion after the summation over $ik_{n}$%
, one obtains 
\begin{equation}
\lim_{iq_{n}\rightarrow \infty }\left( iq_{n}\widetilde{\chi }_{p}\left( 
{\bf q},iq_{n}\right) \right) =-1+n_{\downarrow }+n_{\uparrow }^{0}
\end{equation}
with the help of the normalization Eq.(\ref{Normalisation}) and first moment
sum rule Eq.(\ref{Premier_moment}). This should be compared with the exact
result $\left( -1+n\right) $ found in Eq.(\ref{Coefficient_iqn-1}). The
difference between the two fillings $n_{\downarrow }$ and $n_{\uparrow }^{0}$
is a measure of how much this approximation violates the Pauli principle.

Pursuing the large $iq_{n}$ expansion and using spin rotational invariance
and the sum rules Eqs.(\ref{Normalisation}) to (\ref{Omega_f_A}) on the
single-particle spectral weight, one obtains for the first moment of the
pair spectral weight (off-diagonal $f-$sum rule) 
\begin{eqnarray}
\int \frac{d\omega }{\pi }\omega \chi _{p}^{\prime \prime }\left( {\bf q}%
,\omega \right) &=&\frac{1}{N}\sum_{{\bf k}}\left[ \left( \varepsilon _{{\bf %
k}}+\varepsilon _{-{\bf k+q}}-2\left( \mu -\frac{U}{2}\right) \right) \left(
1-\left\langle n_{{\bf k,}\downarrow }\right\rangle -\left\langle n_{-{\bf %
k+q,}\uparrow }^{0}\right\rangle \right) \right.  \nonumber \\
&&\left. -Un_{\downarrow }-U\left( \left\langle n_{\uparrow }n_{\downarrow
}\right\rangle -n_{\downarrow }n_{\downarrow }\right) +2Un_{\uparrow
}^{0}n_{\downarrow }\right] ~.
\end{eqnarray}
Even with $n_{{\bf k}}$ $=$ $n_{{\bf k}}^{0}$ there are deviations from the
exact result that are linear in $U.$


\begin{references}
\bibitem{NoteSigmaDeGreen}  As usual\cite{KadanoffBaym:1962}, we assume that
the only dependence of ${\bf \Sigma }$ on $\Theta $ is through its
functional dependence on the Green function ${\bf G,}$ Eq.(\ref{DefGreen}).
That ${\bf \Sigma }$ may be expressed as a function of ${\bf G}$ only
requires the same assumptions as those underlying the existence of a
Luttinger-Ward functional\cite{LuttingerWard:1960}. Note also that out of
the four functions entering the matrix ${\bf G}$, only two in principle
suffice to express the dependence on $\Theta $ since the latter has only two
non-zero matrix elements.
\bibitem{NoteTx}  $T_{X}$ denotes the crossover temperature to the
renormalized classical regime. When $T_{X}$ $\ll \mu $, the $SO\left(
3\right) $ symmetry is no-longer apparent and pair fluctuations dominate.
\end{references}
\end{document}